\documentclass[preprint,5p]{elsarticle}
\usepackage{pifont}
\usepackage{natbib}
\usepackage{fleqn}
\usepackage{mathtools}
\usepackage{type1cm}
\usepackage{geometry}
\usepackage{color}
\usepackage{latexsym}
\usepackage{setspace}
\usepackage{hyperref}
\usepackage[all]{hypcap}
\usepackage{graphicx}
\usepackage{epstopdf}
\usepackage{float}
\usepackage{subfig}
\usepackage{multirow}
\usepackage{array}
\usepackage{rotating}
\usepackage{amsmath}
\usepackage{amssymb}
\usepackage{amsthm}
\usepackage[skip=2pt,font=small]{caption}

\usepackage{lineno}

\newcommand{\Deg}{^{\circ}} 
\newcommand{\xmax}{$X_{max}$}
\newcommand{\gcm}{g/cm$^2$}

\begin{document}

\journal{Astroparticle Physics}

\begin{frontmatter}
\setlength{\topmargin}{-35mm}
\setlength{\footskip}{35mm}

\title{Study of Ultra-High Energy Cosmic Ray Composition Using Telescope Array's Middle Drum Detector and Surface Array in Hybrid Mode}

\author[utah]{R.U.~Abbasi} 
\author[saitama]{M.~Abe}
\author[utah]{T.~Abu-Zayyad}  
\author[utah]{M.~Allen} 
\author[utah]{R.~Anderson} 
\author[tokyotech]{R.~Azuma} 
\author[utah]{E.~Barcikowski} 
\author[utah]{J.W.~Belz}
\author[utah]{D.R.~Bergman} 
\author[utah]{S.A.~Blake}
% ...

\address[utah]{High Energy Astrophysics Institute and Department of Physics and Astronomy, University of Utah, Salt Lake City, Utah, USA}
\address[tokyotech]{Graduate School of Science and Engineering, Tokyo Institute of Technology, Meguro, Tokyo, Japan}
% ...
\address[saitama]{The Graduate School of Science and Engineering, Saitama University, Saitama, Saitama, Japan}
% ...

\begin{abstract}Previous measurements of the composition of Ultra-High Energy Cosmic Rays~(UHECRs) made by the High Resolution Fly's Eye~(HiRes) and Pierre Auger Observatory~(PAO) are seemingly contradictory, but utilize different detection methods, as HiRes was a stereo detector and PAO is a hybrid detector. The five year Telescope Array~(TA) Middle Drum hybrid composition measurement is similar in some, but not all, respects in methodology to PAO, and good agreement is evident between data and a light, largely protonic, composition when comparing the measurements to predictions obtained with the QGSJetII-03 and QGSJet-01c models. These models are also in agreement with previous HiRes stereo measurements, confirming the equivalence of the stereo and hybrid methods. The data is incompatible with a pure iron composition, for all models examined, over the available range of energies. The elongation rate and mean values of \xmax\ are in good agreement with Pierre Auger Observatory data. This analysis is presented using two methods: data cuts using simple geometrical variables and a new pattern recognition technique.
\end{abstract}


\end{frontmatter}

\doublespacing

\section{Introduction}

The nature, origin, and propagation of Ultra-High energy Cosmic Rays~(UHECR’s) remains one of the major unsolved questions in particle astrophysics. Recent results from the High Resolution Fly's Eye~(HiRes),  Pierre Auger Observatory~(PAO), and Telescope Array~(TA) experiments~(\cite{FirstGZK2008},\cite{PAOGZK},\cite{AbuZayyad201316}) have reliably determined that the spectrum of these cosmic rays terminate near 60~EeV, consistent with predictions of a cutoff (so-called GZK cutoff~\cite{Greisen1966},~\cite{Zatsepin1966}) due to the onset of inelastic interactions of protons with the primordial 2.7 K black body radiation. Such a cut-off implies that the sources of the highest energy cosmic rays must lie relatively nearby~($\lesssim$~100~Mpc).

The nature of the composition of these cosmic rays is critical in determining whether this is in fact the mechanism, since such a cutoff could be mimicked in a variety of other ways~\cite{Allard200761},\cite{Berezinsky2014120},\cite{Kampert2014318}. The composition of UHECR also has a major impact on predictions of the minimal extragalactic neutrino flux, as well as expectations for determining sources of UHECR, by searching for anisotropy. If cosmic ray composition is light, and our understanding of extragalactic and galactic magnetic fields are not far from being correct, then sources within the 100~Mpc GZK radius should become evident as anisotropic enhancements. The situation becomes much less encouraging if the composition is heavy.

Because of the low flux of UHECR it is unfeasible to study them by direct detection. Instead, we determine the longitudinal shape of the air-shower of particles produced by the interaction of the primary cosmic ray in the atmosphere, using the air-fluorescence technique pioneered by the Fly's Eye, and HiRes experiments~\cite{Baltrusaitis1985},\cite{AbuZayyad2000253}. The extensive air shower~(EAS) reaches a maximum in particle density at a point in the atmosphere where the mean energy of the secondaries drops below the critical energy. The distribution of the depth of this maximum (\xmax) is sensitive to the nature of the primary composition. Heavy nuclei will interact higher in the atmosphere, and have smaller fluctuations in shower development, while protons will interact more deeply, and have larger \xmax\ fluctuations. 

Comparison of HiRes, TA, and PAO results is complicated by the different analysis approaches of the experiments. In the case of HiRes, loose quality cuts to ensure good resolution and minimal energy dependent biases in \xmax\ are applied to data and Monte Carlo (MC) simulated proton, and iron events in an identical manner. It is seen that all geometrical variables such as zenith, impact parameter, etc. are in excellent agreement between data and MC, at all relevant energies. Residual acceptance, and reconstruction bias, is dealt with by comparing the well modeled final \xmax\ and energy MC distributions with data. Data are not shifted to take into account biases, but model predictions include all detector effects, and biases. A similar approach is used in this paper. In the case of PAO, tight cuts are devised to remove acceptance and reconstruction bias, so that the data can be compared directly to the thrown, unbiased, simulated data. Both approaches should give consistent results; if systematics and methodology are well understood. Because stereo (HiRes), and hybrid detection (this paper), have different acceptances in zenith, impact parameter, core position, and \xmax\ with concomitant differences in the degree of atmospheric attenuation, and detector coverage of shower profile, as well as using completely different variables to reconstruct shower geometry (e.g. geometry based on intersection of planes versus fits to tube timing) -- the detector and reconstructed MC are tested in very different ways. A consistent result between stereo, and hybrid, methods would further validate the efficacy of the HiRes/TA approach in taking acceptance, and reconstruction biases, into account.

 Good control of geometric reconstruction, sophisticated modeling of the fluorescence detector response, good calibration of the detector system, and continuous modeling of the atmosphere, are thus essential. Earlier results from the Fly's Eye, and HiRes experiment, using a purely air-fluorescence technique and stereo observation of each EAS, indicated a predominantly light composition of cosmic rays~\cite{Abbasi2005},\cite{Abbasi2010a}. More recently, PAO in the Southern hemisphere presented results obtained from fluorescence detectors operated in hybrid mode in conjunction with a surface array of Cherenkov water tanks. Their conclusions were that for UHECR  ``10$^{18}$ to 10$^{18.5}$~eV... the shape of the \xmax\ distribution is compatible with there being a substantial fraction of protons..."~\cite{PhysRevLett.109.062002} and ``a gradual increase of the average mass of cosmic rays with energy up to 59~EeV "~\cite{PhysRevLett.104.091101} was evident.

\begin{figure}[t!]
  \centerline{
   \includegraphics[width=1\linewidth]{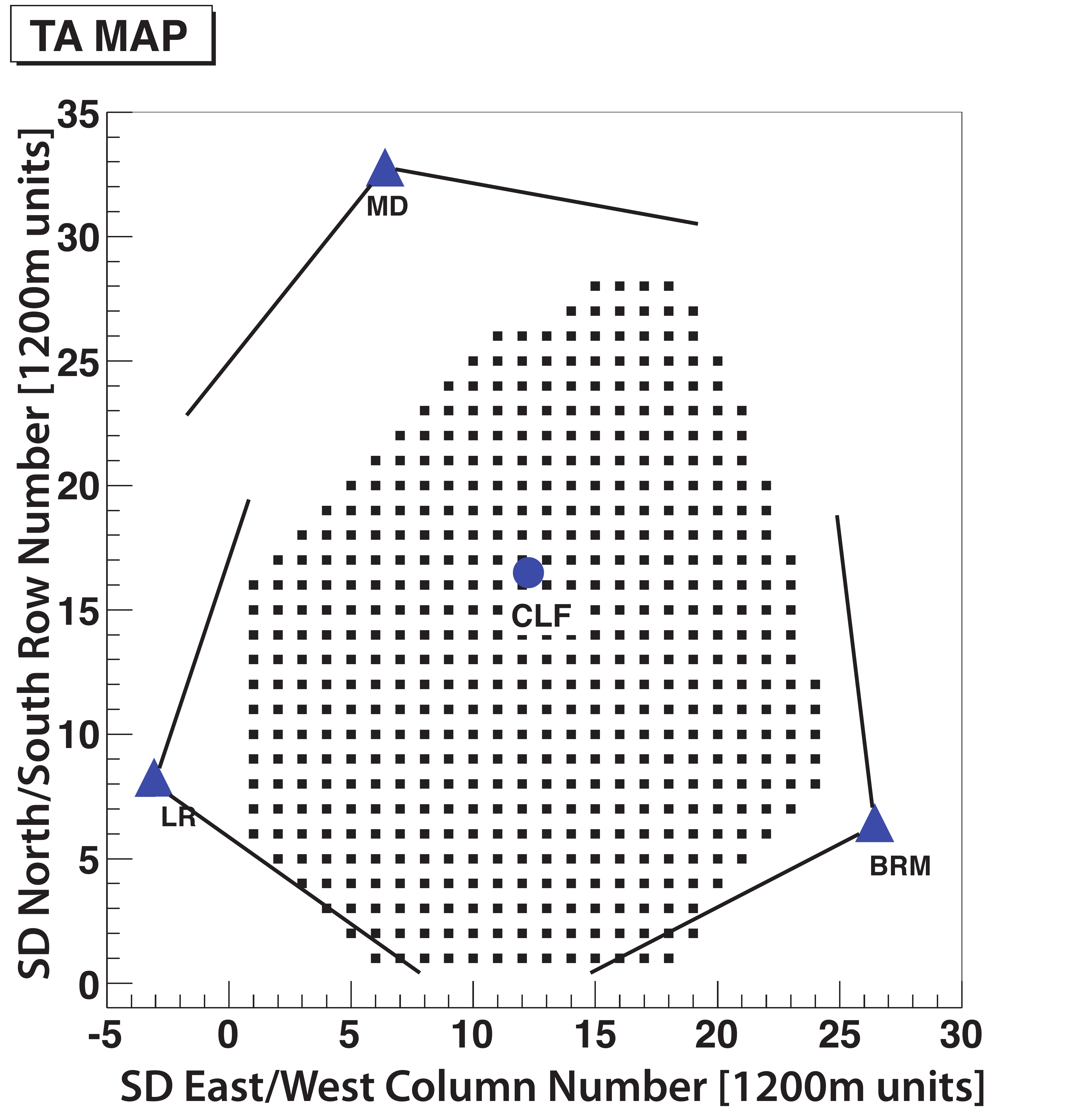}
   }
   \caption[The layout of the Telescope Array experiment.]{The layout of the Telescope Array experiment. Filled black squares indicate the locations of the 507~scintillation counters that comprise the Surface Detector~(SD) array. Triangles mark the three fluorescence detection sites at the periphery of the SD array. The Central Laser Facility~(CLF), shown by the circle, is placed equidistant from the three fluorescence detection sites to provide atmospheric monitoring, and cross-calibration.} 
  \label{fig:map}
\end{figure}

The composition analysis discussed in this paper uses five years of hybrid measurements from the Surface Scintillation Detector~(SD) array in conjunction with the Fluorescence Detectors~(FDs) at the Middle Drum~(MD) site, at the northernmost end of the TA experiment. This site is unique in that the equipment consists of 14~refurbished telescopes from the HiRes-1 site of the previous HiRes experiment~\cite{Rodriguez2011}. Figure~\ref{fig:map} shows the layout of the Telescope Array experiment. The 507~SDs (black squares) are located in a~1.2~km square grid surrounded by three FD sites (blue triangles) that overlook the SD array~\cite{Allen2012}.

A composition measurement of UHECRs requires an accurate measurement of the longitudinal profile of the cosmic ray showers generated by the particle. The FDs measure the fluorescence light emitted by the excited atmospheric molecules, due to the charged particles in the developing shower. Using the known fluorescence yield, one can calculate the energy of the shower from the fluorescence light. This measurement, however, is only robust if the location, and direction, of the shower are also calculated accurately. Using the FDs in monocular mode, to calculate composition, has the disadvantage of potentially large systematic errors in geometry. Hybrid detection allows us to take advantage of the SD array, which samples the lateral distribution of the particle showers as they hit the ground. The local densities of the particles, along with the arrival times, are measured, and used to calculate the core location, and the geometry of the shower. 

Due to the large event-to-event fluctuation inherent in air shower development, indirect detection techniques (such as fluorescence) are unable to determine the mass of any individual primary cosmic ray. Instead, the characteristics of the longitudinal development of events are used to give a statistical measure of the composition. Specifically, the distribution of events is used to distinguish between showers produced by light particles, and those produced by heavy particles.

This paper approaches the analysis of the hybrid \xmax\ data in two different ways. Since, the MD detector is essentially identical to the HiRes detectors (except for a more limited azimuthal field of view), we first examine the \xmax\ distributions using simple cuts on geometrical variables, similar to those previously used in the HiRes analysis. Due to the 10~km distance between the MD FD and the edge of the SD array, which causes a low average photon count at low energies, the MD hybrid data \xmax\ resolution is a strong function of energy. Because of this effect, we also developed a new technique using pattern recognition to improve the resolution, and to minimize its energy dependence. This analysis forms the second part of the paper starting at Section~\ref{sec:PatternRecognition}.  

\section{Hybrid Event Reconstruction}
The MD hybrid composition analysis begins with event reconstruction. This study uses programs that were created to reconstruct events in monocular mode by the FDs, and stand-alone SDs. These reconstruction steps are performed independently on the initial data from each detector, and the results are combined into a hybrid analysis. Specifically, the particle density, and timing information from the SDs, are combined with the longitudinal profile from the FDs, to generate a hybrid picture from which to calculate the \xmax\ parameter.

\subsection{SD Reconstruction}
Each SD consists of two layers of 3~m$^{2}$~scintillator. As particles interact with the material the light emitted is directed along wavelength shifting fibers to one of two Photo-Multiplier Tubes (PMTs) (one for each layer)~\cite{ICRR2006}. The analog signal from the PMTs is digitized by a Flash Analog to Digital Converter (FADC) every 20~ns. A Minimum Ionizing Particle (MIP), or the average signal from a single cosmic ray muon, is used for calibration. The FADC traces are scanned and any pulse that exceeds 0.3~MIPs is saved with a time stamp. An event is triggered when 3~adjacent counters have signals greater than 3~MIPs within an 8~$\mu$s window. At that point, all the signals greater than 0.3~MIPs within a 32~$\mu$s coincidence window are assigned to that event~\cite{Ivanov2012}.

Individual SD counters, not contiguously connected on the array, or with signals not contiguous in time, are removed from the event. An initial geometry calculation of the shower is then performed using the trigger times of each SD counter in the event. Finally, a lateral distribution function is used to fit the particle densities at a lateral distance, perpendicular to the shower core location, generating a more accurate geometrical reconstruction of the shower. This geometry is used for the hybrid analysis. The SD design and event reconstruction techniques are described in more detail in~\cite{Ivanov2012} and~\cite{AbuZayyad201287}.

\subsection{FD Reconstruction}
Each of the 14~telescopes at the MD site consists of a 5.1~m$^2$ spherical mirror that collects light from the cosmic ray shower and images it onto a camera. The camera consists of 256~PMTs behind a UV band-pass filter that is used to remove extraneous light, thereby improving the signal to noise ratio. The PMTs are positioned in subclusters consisting of 16~PMTs in a 4$\times$4~unit. Individual PMT thresholds are modified continuously to keep their trigger rate at a constant 200~Hz. In order to trigger a subcluster, at least three tubes, with two being adjacent, must all trigger within a 25~$\mu$s window. In order to trigger an event, two subclusters must trigger within a 25~$\mu$s window. The data from triggered telescopes are combined using the timing information and events across telescopes are saved as one event.

Data from a single event is processed using a Rayleigh filter. The filter examines the path of the tubes to determine if the event is triggered by noise, or an actual cosmic ray event. Finally, the Shower Detector Plane~(SDP) is calculated from the pointing directions of the PMTs. This is done by treating the SDP as a line source. The fit uses $\chi^2$ minimization of Equation \ref{eq:SDPchi2}. 

\begin{equation}
\alpha^2 = \sum_i \frac{(\mathbf{\hat{n}} \cdot \mathbf{\hat{n_{\emph{i}}}})^2 \cdot w_i}{\sigma_i^2}
\label{eq:SDPchi2}
\end{equation}

\noindent In this equation, the SDP normal vector is represented by $\mathbf{\hat{n}}$, and the viewing direction of the triggered tube $i$ is $\mathbf{\hat{n_i}}$. The signal in tube $i$, or number of photoelectrons, is $w_i$.  The angular uncertainty of each tube, $\sigma_i$, is set to 1$\Deg$. Reconstruction using the MD data in monocular mode is described in detail in~\cite{Rodriguez2011}.

\subsection{Hybrid Reconstruction}
The hybrid reconstruction method for the composition analysis was previously used for an energy spectrum analysis, described in~\cite{Allen2012}. Once the initial separate reconstructions have been done for SD and FD events the two are combined into one hybrid set using the timing information. The time that the shower core passes the SD plane can be calculated independently from SD and FD measurements, and any events that occur within 2~$\mu$s of each other are assumed to be one hybrid event.

Each hybrid event is processed taking into account timing information, from both the FD and SD, as well as the position of the shower core on the ground measured by the SD. A $\chi^2$~minimization is done taking into account all of these pieces.

The calculation of the shower geometry in the Shower Detector Plane~(SDP) depends on the timing and pointing directions of the PMTs. The relationship between the event geometry, and the timing in the SDP, is described by Equation \ref{eq:TVSA}.
\begin{equation}
t_{i} = T_{R_P} + \frac{R_{P}}{c} tan \left( \frac{\pi - \psi - \chi_{i} }{2} \right)
\label{eq:TVSA} 
\end{equation}

\noindent Here, $t_i$ represents the time that the $i^{th}$ tube triggered. $T_{R_P}$ is the time of the shower at the impact parameter, ($R_P$), and the tube viewing angle is represented by $\chi_{i}$. Using this relationship, the angle of the shower, $\psi$, within the SDP can be calculated. More PMTs lead to a more accurate measurement. The hybrid analysis takes advantage of this possibility by adding the SDs as virtual ``PMTs" to the calculation. This is done using the trigger time of the SDs, taking into account the distance for light to travel from the SD to a hypothetical PMT at the MD detector. Once all the SDs are included in the calculation, a more accurate measurement of the geometry is obtained, and is used in the next stage of the analysis. 

Next, the shower core position calculated from the SD analysis is used to constrain the hybrid analysis. The SDP normal vector $\mathbf{\hat{n}}$ (from Equation \ref{eq:SDPchi2}) that is fit during the MD FD analysis, is combined with the SD array information, and a $\chi^2$ minimization is done. The fit takes into account the timing, as well as the core location. Three parameters are varied to find the minimum: $\psi$, the angle of the shower within the SDP; $R_{P}$, the impact parameter; and $T_{Rp}$, the time that the shower is closest to the Middle Drum FD detector. These three parameters, along with the SDP, completely describe the geometry of the shower. More detailed information can be found in~\cite{Allen2012}.

The final step, of the hybrid composition analysis, is to reconstruct the shower profile to find the \xmax\ of the shower. Each PMTs view of the shower is first converted into a shower depth, in \gcm, and the relationship between the slant depth and the signal size for each PMT is compared to a library of Monte Carlo~(MC) simulated showers generated by CORSIKA~\cite{Heck1998}. 

The MC showers are parametrically calculated, using Poisson statistics. The initial energy of a cosmic ray particle is given, and the number of particles, $N_{e}$, at any slant depth, $x$, is calculated from the Gaisser-Hillas parameterization~\cite{gaisserhillas} given in Equation~\ref{eq:GHf}.
\begin{equation}
N_{e}\left(x\right)  = N_{max} \times \left[\frac{x-X_{0}}{X_{max}-X_{0}}\right]^{\frac{X_{max}-X_{0}}{\lambda}} exp\left(\frac{X_{max}-x}{\lambda}\right) \, ,
\label{eq:GHf}
\end{equation}

\noindent $N_{max}$ is the number of particles at the shower maximum and \xmax\ is the slant depth at the shower maximum. The signal size of a PMT, at any slant depth, is then calculated given the initial energy of the MC shower. The signal size of each PMT in a hybrid event is compared with the predicted signal size of the MC shower at the same slant depth. Then, the $\chi^2$ function is calculated, and the shower is matched with the MC shower that it best represents. The value of \xmax\ and energy is then taken from the MC shower. The SD energy is renormalized to the FD spectrum, as has been done in the TA SD spectrum analysis. Details about energy calibration and corrections for missing energy can be found in ~\cite{Allen2012}.

\section{Geometry Cut Event Selection}
After a study of simulated showers, cuts were made on the data to improve the quality of the reconstruction of the shower parameters. With these cuts the data and MC distributions of various geometrical quantities such as zenith, $R_P$, track length, etc. are found to be in good agreement with each other. These are the same cuts that were used in the spectrum analysis, and are listed below~\cite{Allen2012}. 
\begin{enumerate}
\item Weather cut: To ensure that clouds do not obscure, or limit, the field of view of the FD; only events that occur on clear nights with no visible clouds are included in the sample.
\item Failmode: Events that failed the profile reconstruction are removed from the set. 
\item Zenith angle ($>$56$\Deg$): Events with zenith angles greater than 60$\Deg$ cannot presently be reconstructed reliably using the surface detector technique. Therefore, the Monte Carlo for this analysis does not contain simulated showers with zenith angle greater than 60$\Deg$, and no determinations can be made about them.
\item Hybrid/Surface Detector Core Position (difference $>$1200~m): Since the events are time-matched, it is conceivable that two independent events (one SD event and one MD event) may be combined due to their proximity in time. The core location of the shower at the ground, calculated using only the SDs (see \cite{Ivanov2012} for details of SD reconstruction), is compared to the position calculated using the hybrid analysis, to ensure that the MD event and the SD event are the same event, so that only true hybrid events are kept. 
\item Border Cut ($<$100~m): The border cut uses the hybrid core location to determine how close the shower falls to the edge of the SD array. Each shower core is required to be inside the border of the array. The majority of showers with calculated core locations that fall at, or outside, the border of the array were difficult to reconstruct.
\item Track Length ($<$8.0$\Deg$): Events with shorter track lengths have less information, and therefore, provide a less accurate reconstruction.   
\item \xmax\ not ``Bracketed": Events which are reconstructed with the depth of \xmax\ outside of the field of view of the detector camera (3-31$\Deg$ elevation) are removed. The energy and composition are determined more accurately if \xmax\ is seen.
\item Energy~($<$10$^{18.2}$~eV): Due to the distance between the FD, and the SD, events with energies below this value have poor trigger efficiency. The reconstruction accuracy is also poor due to low FD photon yields.
\end{enumerate}

The number of events which pass the weather and failmode cuts is~1916. The number of events which pass all cuts is~843, corresponding to 44\% of the total good weather reconstructed events, in the five year time period between May 2008 and May 2013.

\section{Composition Results with Geometry Cuts}
The \xmax\ parameter helps distinguish between light, proton-like, and heavy, iron-like showers in two ways: (1) $<$\xmax $>$, the average \xmax\ value: simulations shows that proton-induced showers tend to penetrate further into the atmosphere, and develop later, resulting in a larger $<$\xmax$>$ value than iron-induced showers of the same energy. Heavy particles tend to interact sooner, and produce a much larger multiplicity of secondaries on the first interaction, resulting in a smaller $<$\xmax$>$ value. The $<$\xmax$>$ from the data are compared to a set of Monte Carlo~(MC) events, using proton primary particles, and a set using iron nuclei primary particles, to determine which set best describes the data. Actual cosmic rays may have intermediate nuclei, and/or a mixture of heavy and light particles. Due to statistical limitations, this analysis only deals with the two composition extremes. (2) The distribution of the \xmax\ value: proton-induced showers have a smaller multiplicity of particles in the first interaction, which results in greater fluctuations, and a wider distribution of \xmax\ values (larger RMS),  while iron-like showers produce a narrower distribution.

\begin{figure}[tbp!]
  \centerline{
   \includegraphics[width=0.8\linewidth]{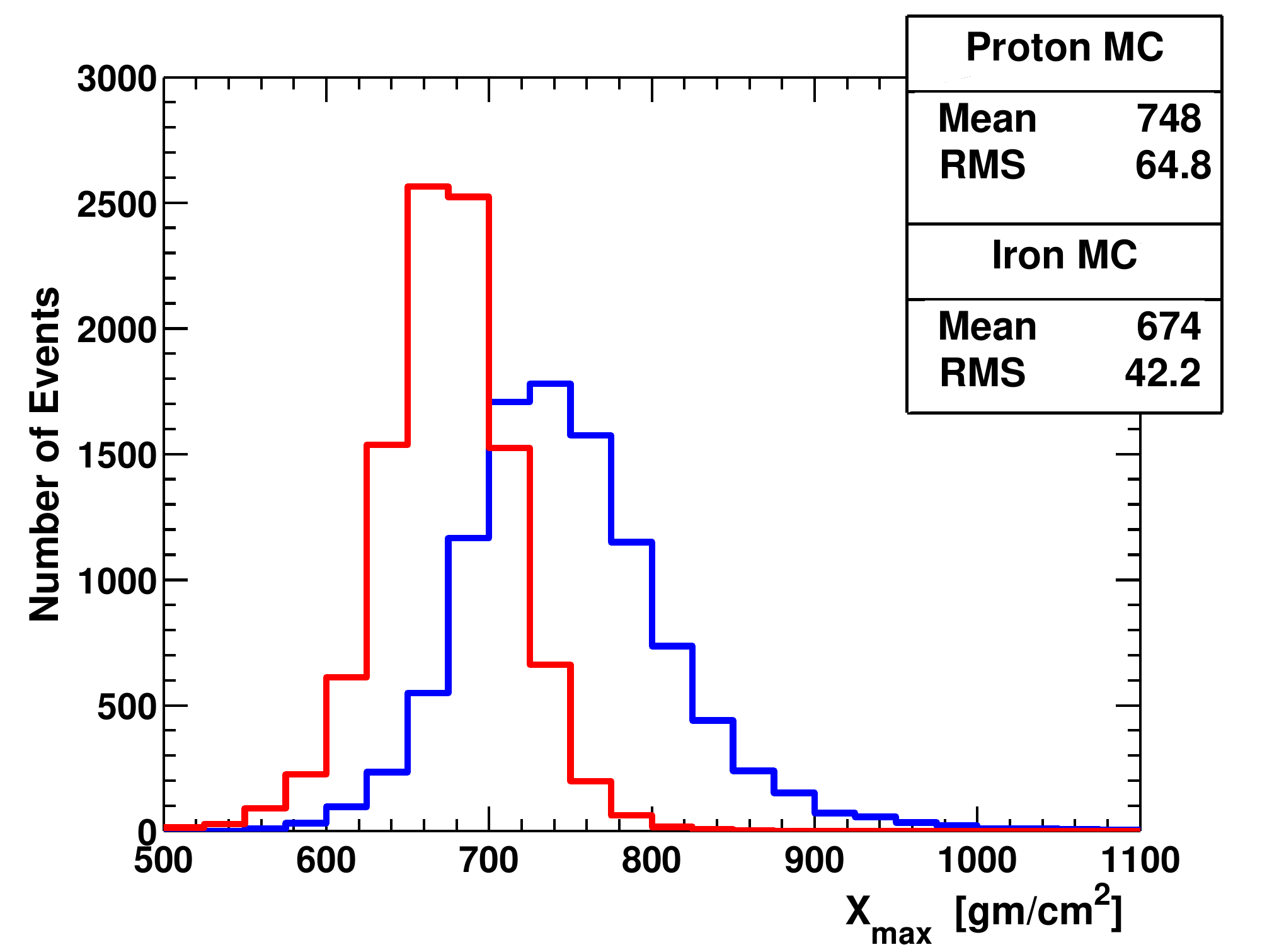}
}
\caption[The distribution of the shower maximum for the proton/iron Monte Carlo set]{The distribution of the shower maximum for the proton Monte Carlo set (blue), and the iron Monte Carlo set (red), using QGSJETII-03. These distributions show all events with reconstructed energies E~$>$10$^{18.2}$~eV. Note that the mean \xmax\ of the proton data is deeper and that the width of the proton distribution is significantly wider. The significant overlap between the two distributions make an event-by-event identification impossible.}
\label{fig:MCXmaxDist}
\end{figure}

The composition study in this analysis requires the use of two Monte Carlo sets: one thrown with iron nuclei, and one thrown with protons. The two sets were thrown in the same manner, the only difference being the primary particle. Several hadronic model simulations are used. In what follows we compare to QGSJET-II-03~\cite{Ostapchenko2006}. Other model generators are discussed in Section~\ref{sec:TAPAO}. The proton MC set contained 21,649~events which would have triggered the detector in hybrid mode. After quality cuts 10,070~events remained. The iron MC set contained 24,295~events which would have triggered the detector. With cuts, 11,335~events were kept. Figure \ref{fig:MCXmaxDist} shows the reconstructed \xmax\ distributions, for energies $>$10$^{18.2}$~eV of the proton and iron MC sets. The mean \xmax\ value of the proton set is 748~\gcm, significantly higher than the mean of the iron set, 674~\gcm. 

Note that, though the $<$\xmax $>$ of the proton set is deeper than iron, the width of the proton distribution is significantly wider.  The fact that there is significant overlap between the two distributions make an event-by-event composition identification impossible.

\begin{figure}[tbp!]
\centerline{	
\includegraphics[width=0.8\linewidth]{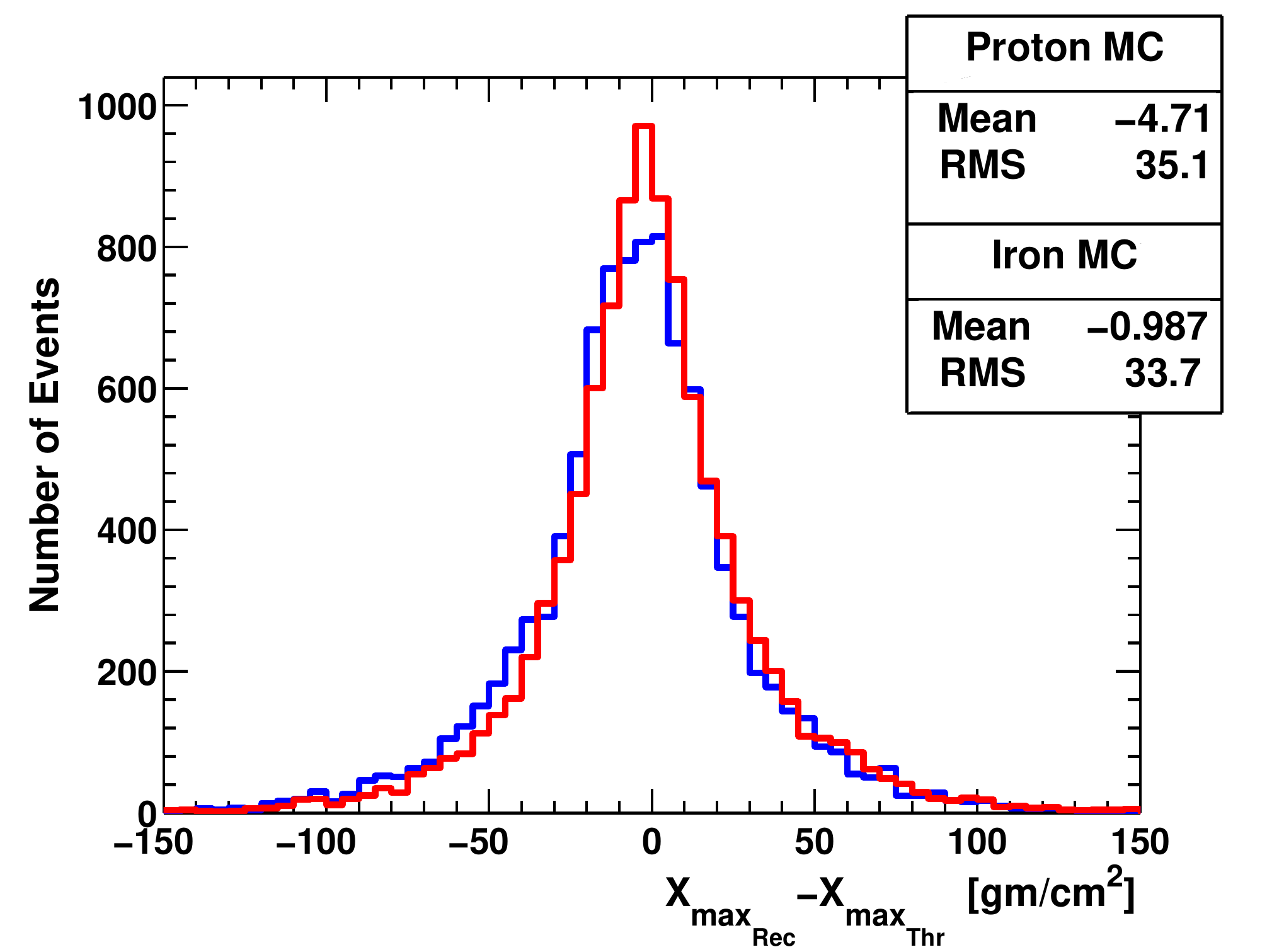}
}
\caption{The MD hybrid \xmax\ resolution using geometry cuts, above E~$>$10$^{18.2}$~eV, for the QGSJETII-03 Monte Carlo sets: shown is the difference between the reconstructed \xmax\ values, and the thrown \xmax\ values, for proton induced showers (blue), and iron induced showers (red).} 
\label{fig:MCResXmax}
\end{figure}

Figure \ref{fig:MCResXmax} shows the resolution of the reconstructed \xmax\ (RMS of the difference between reconstructed and simulated values) above E~$>$10$^{18.2}$~eV, for the proton and iron Monte Carlo respectively. The overall resolution of the proton set is 35.1~\gcm, with a bias of -4.7~\gcm. Resolution of the iron set is comparable, with a width of 33.7~\gcm\ and a bias of -0.99~\gcm. The pattern recognition method for further improving the resolution is discussed in Section~\ref{sec:PatternRecognition}.

\begin{figure}[tbp!]
\centering
\subfloat[Subfigure 1][]{
\includegraphics[width=0.8\linewidth]{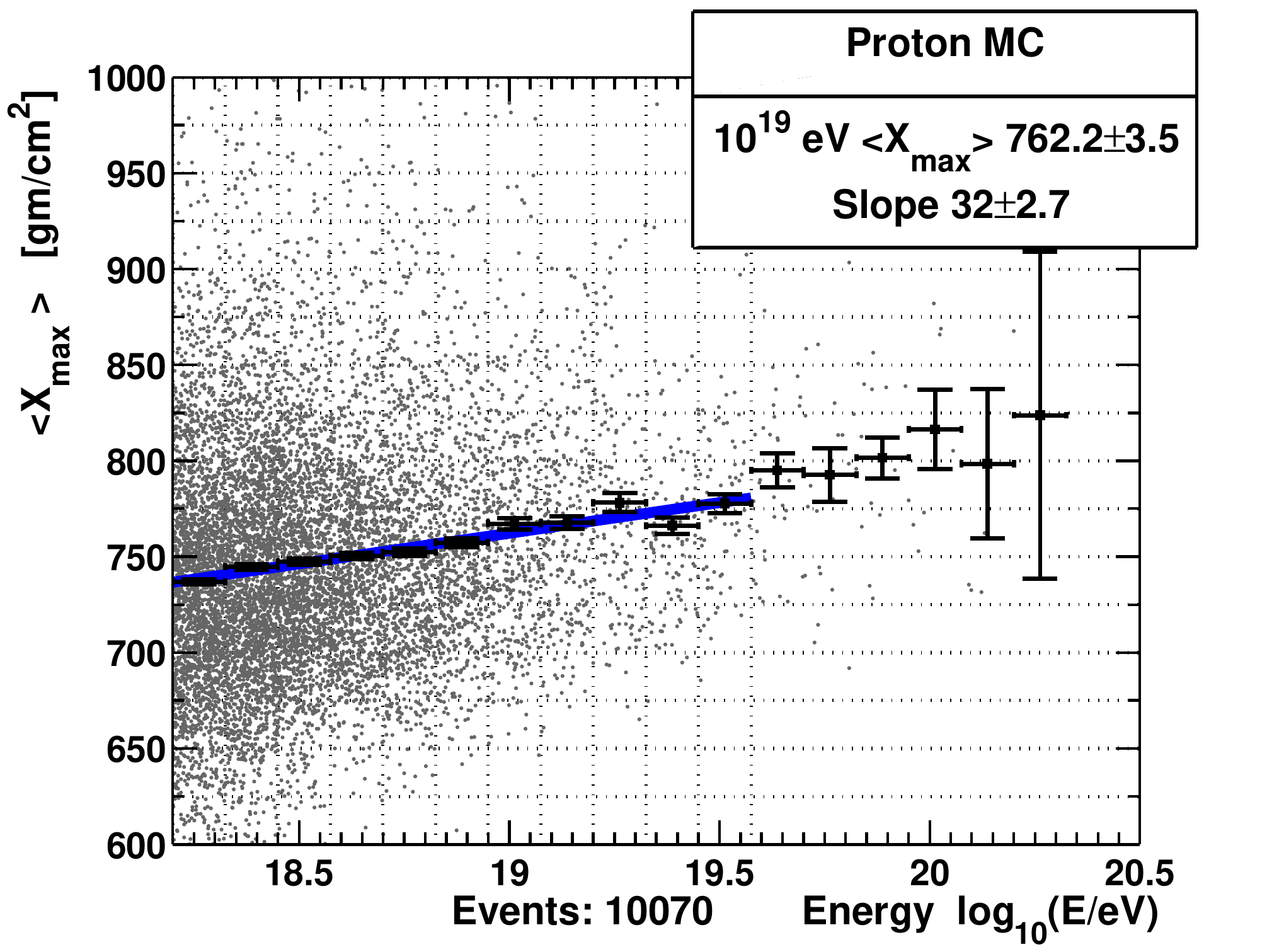}
\label{fig:ERateProton}}
\qquad
\subfloat[Subfigure 2] [] {
\includegraphics[width=0.8\linewidth]{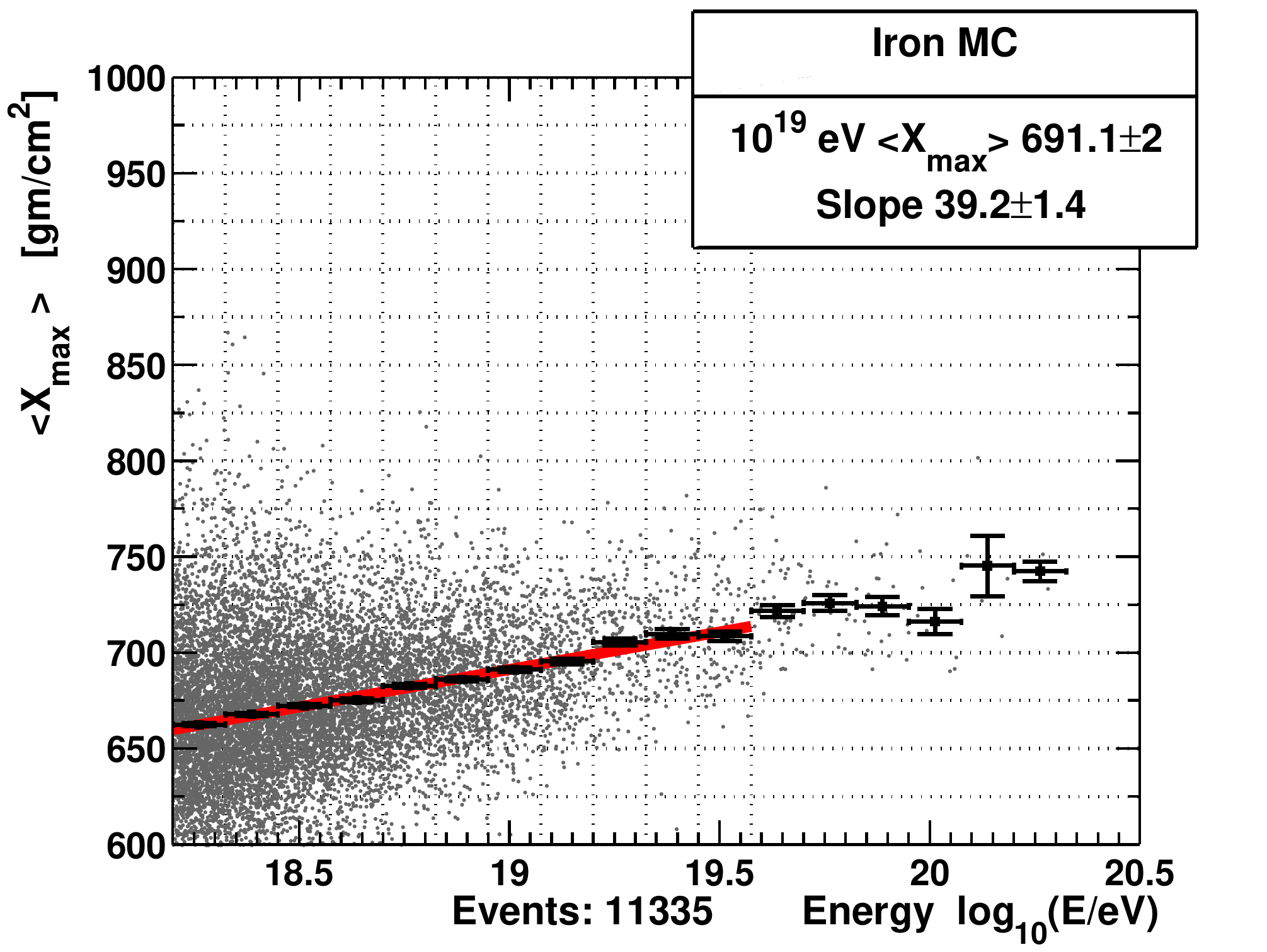}
\label{fig:ERateIron}}
\caption[Middle Drum hybrid MC \xmax\ distribution scatter plot: the $<$\xmax$>$ values are plotted as a function of energy.]{Middle Drum hybrid MC \xmax\ distribution scatter plot: (a) proton and (b) iron induced QGSJETII-03 MC are shown, along with the apparent elongation rate, or slope, of the fit. \xmax\ values (grey points) are plotted as a function of energy. The black data points with error bars represent the $<$\xmax$>$ values, in bins that are plotted as a function of bin energy. The solid lines are the fits to the $<$\xmax$>$ values, up to the energy for which there is low statistics in the data.} 
\label{fig:MCERate}
\end{figure}

Figures \ref{fig:ERateProton} and  \ref{fig:ERateIron} show scatter plots of \xmax\ values, as a function of their shower energy, for the proton and iron Monte Carlo respectively. The points with error bars represent the $<$\xmax$>$ and error on the mean in each energy bin. The line is a fit to the $<$\xmax$>$ values, up to the point where there is low statistics in the data. The slope of the line, or elongation rate, is 32~\gcm/energy~decade for the proton, and 39~\gcm/energy~decade for iron. The measured elongation rate takes into account the detector and reconstruction bias that is included in the Monte Carlo, and therefore does not represent the true elongation rate of cosmic ray particles. The model dependence of the elongation rate is discussed in Section~\ref{sec:TAPAO}.

\begin{figure}[tbp!]
\centerline{
\includegraphics[width=0.9\linewidth]{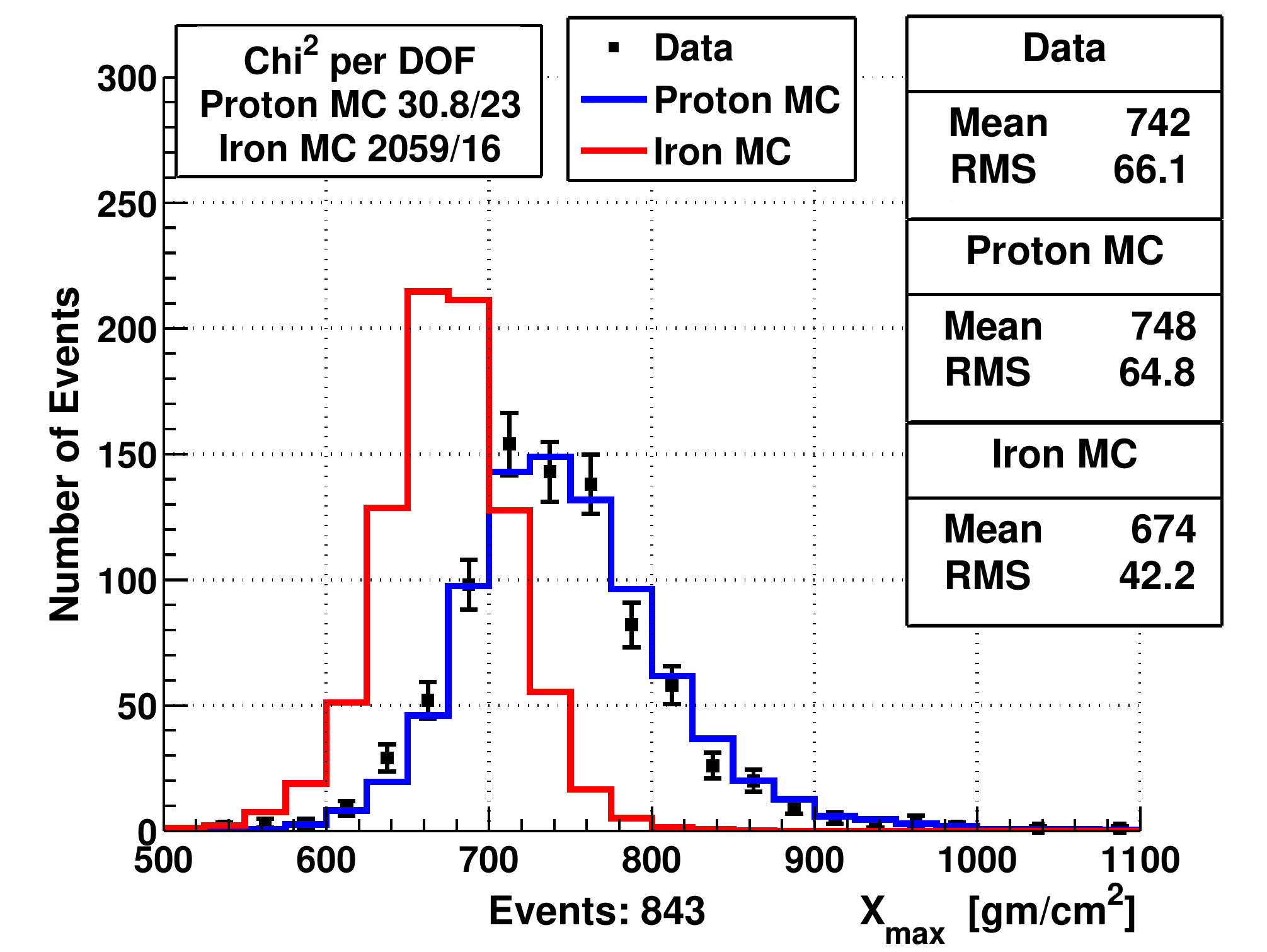}
}
\caption[Data/MC comparisons of the shower maximum (\xmax): the distribution of measurements, for E~$>$10$^{18.2}$~eV, is shown for the data (black points with error bars) with the proton MC (blue), and the iron MC (red) histograms.]{Data/MC comparisons of the shower maximum (\xmax), for E~$>$10$^{18.2}$~eV: the distribution of measurements are shown for data (black points with error bars), QGSJETII-03 proton MC (blue), and iron MC (red) histograms. The MC has been normalized to the area of the data. A binned maximum likelihood estimated chi-squared test value shows agreement between data and proton MC.} 
\label{fig:DTMCProtIron}
\end{figure}

Figure \ref{fig:DTMCProtIron} shows the overall data/MC comparison of the \xmax\ distribution for E~$>$10$^{18.2}$~eV, for the proton and iron Monte Carlo. The mean \xmax\ value of the data set is 743~\gcm. The binned maximum likelihood estimated chi-squared test value was calculated to compare the distributions~\cite{Baker1984}. Note that the proton distribution is in much better agreement than the iron distribution. This confirms previous findings by the HiRes and PAO collaborations.

\begin{figure}[t!]
  \centerline{
   \includegraphics[width=0.9\linewidth]{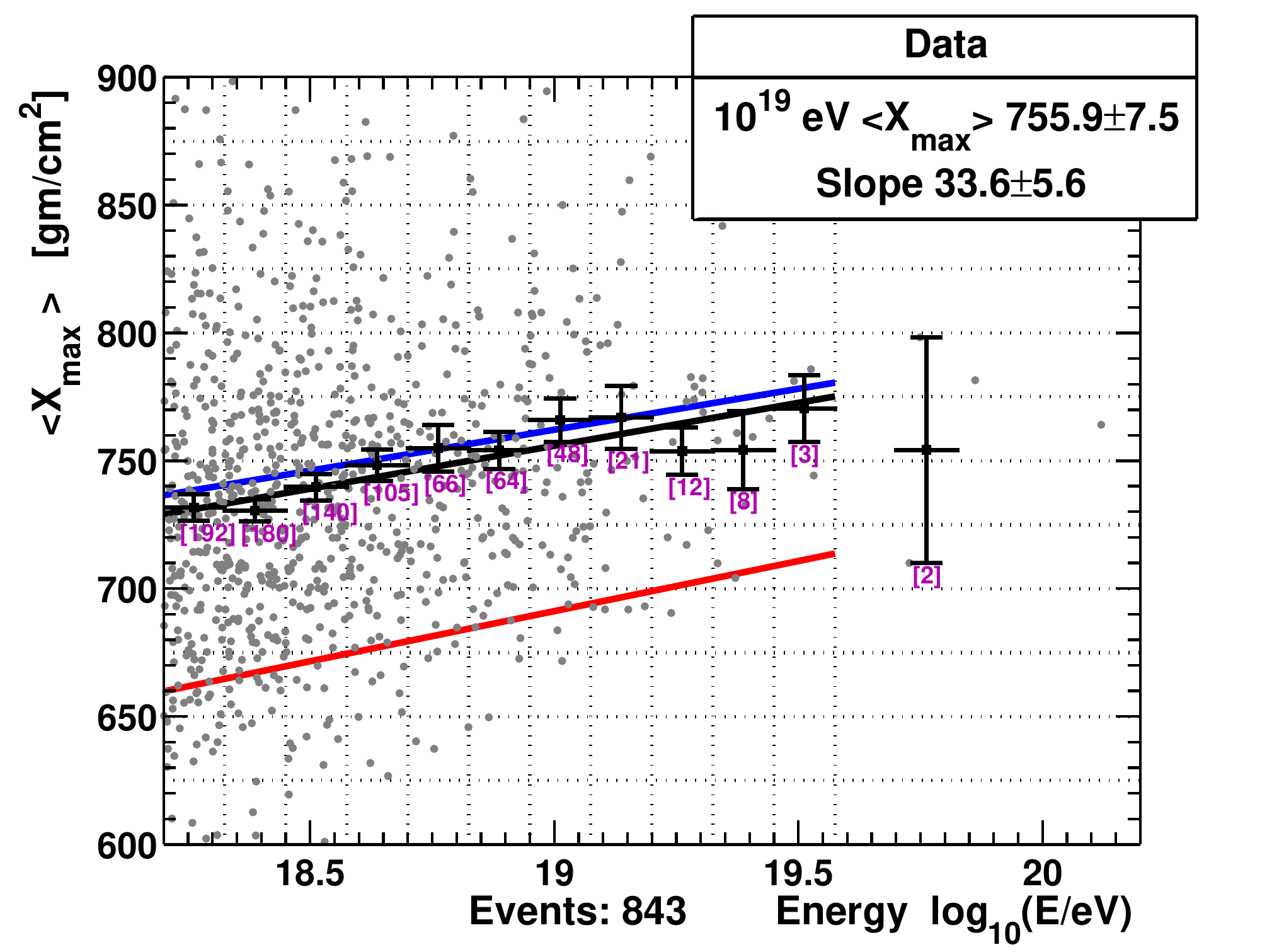}
   }
   \caption[Middle Drum hybrid composition result using geometry cuts: the $<$\xmax$>$ values for each data event are plotted as a function of energy overlaid with the proton and iron ``rails". ]{The five year Middle Drum hybrid composition result using geometry cuts: \xmax\ values (grey points) for each data event are plotted as a function of energy, overlaid are QGSJETII-03 proton (blue) and iron (red) MC ``rails" from Figure~\ref{fig:MCERate}. Black data points with error bars represent the data $<$\xmax$>$ values, in 12~energy bins (of width $\log_{10} (E/eV) = 0.125$), that are plotted as a function of bin energy. The black rail is a fit to these binned values. All rails are fitted up to the energy for which the data has low statistics. The number of events in each bin is listed below the error bars. The scale is chosen for emphasis on elongation, this cuts 24 events from the scatter plot.} 
  \label{fig:ERateTotal}
\end{figure}

Figure \ref{fig:ERateTotal} shows a scatter plot of the data \xmax\ values, as a function of their energy, compared with the proton and iron Monte Carlo. The proton and iron fitted lines, or ``rails", are taken from plots \ref{fig:ERateProton} and \ref{fig:ERateIron}, indicating the $<$\xmax$>$ values of the MC showers. The data clearly agree more closely with the proton rail than the iron rail. Recall that the proton and iron MC sets have used the same reconstruction programs, as well as the same cuts, as the data set.

We conclude that hybrid reconstruction, using a similar analysis to what was used in HiRes, leads to similar conclusions. The results are independent of the stereo, or hybrid, technique.

While an overall look, at the \xmax\ distributions, can give some insight into the composition of the primary particles in the data, it does not give a complete picture. The cosmic ray particle composition could be energy dependent. Therefore, a study of how this distribution evolves with energy is suitable, and the distribution of the \xmax\ parameter of the data, compared to both MC sets, are examined in smaller energy ranges. However, the current set of cuts produce an energy dependent \xmax\ resolution (Figure~\ref{fig:monicaresolution}). Much of the resolution energy dependence comes from the increasing number of events, at lower energies, that do not show a pronounced shower maximum in the detector field of view. In particular, the PAO results indicate an energy dependent narrowing of the \xmax\ distribution~\cite{PhysRevLett.104.091101}. It is therefore important to reduce the resolution energy dependence, over as large an energy range as possible, to improve the reliability of our conclusions. As described in Section 5, simple chi-squared cuts on the G-H fits are not sufficient to reject most low energy events that have poorly defined \xmax. 

To this end, we have developed a pattern recognition program that selects events that have a clear rise and fall before, and after, the putative shower maximum. Figure \ref{fig:optresolution} shows the improvement in the \xmax\ resolution energy dependence from imposing this selection.

\section{Description of Pattern Recognition Method} \label{sec:PatternRecognition}

\begin{figure}[tbp!]
  \centerline{
   \includegraphics[width=0.9\linewidth]{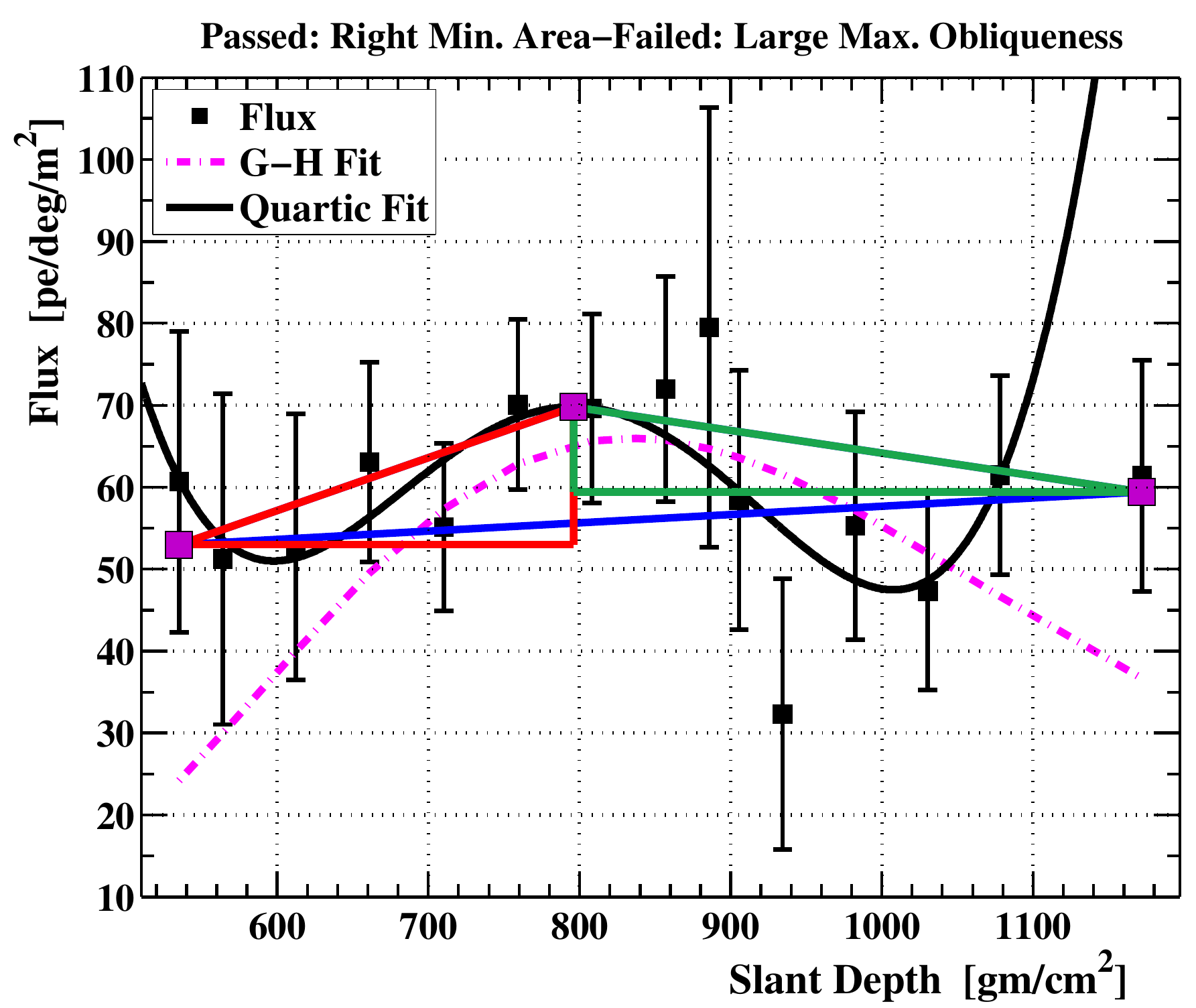}
   }
   \caption{An event which passed the right triangle area test, but failed the large triangle obliqueness test. Bins with large errors have been removed for display purposes.} 
  \label{fig:Event241}
\end{figure}

Only events which have a clear rise and fall in FD photon signal flux versus atmospheric depth contain information on \xmax\ that can be reliably reconstructed. These events will have the best \xmax\ resolution. At lower energies, showers are only sufficiently bright to trigger the detector near shower maximum, resulting in a relatively flat profile with little curvature (See Figure~\ref{fig:Event241}, for example). Events with shower maximum either above, or below, the field of view of the detector will result in a monotonically increasing, or decreasing, profile. The position of shower maximum must then be extrapolated, which leads to additional errors, and a systematic dependence on the assumed form of the fitting function.  While the effect of these events can be reduced by fitting a Gaisser-Hillas~(GH) function~\cite{gaisserhillas} to the profile and demanding that the resulting position of \xmax\ be in the field of view, the issue of a flat profile is not easily dealt with in this way. Lower energy events have relatively large statistical errors in signal bins, and a simple chi-squared goodness of fit test, to the GH profile, will not give a good discrimination. In fact, many quite flat profile events produce a good chi-square fit. A different approach is needed to remove these events.

A simple pattern recognition method has been created which rejects flat events, or events which only have a rise, or a fall, in signal magnitude, but not both. This significantly improves the overall \xmax\ resolution, and the energy dependence of the resolution.

\begin{figure}[tbp!]
  \centerline{
   \includegraphics[width=0.9\linewidth]{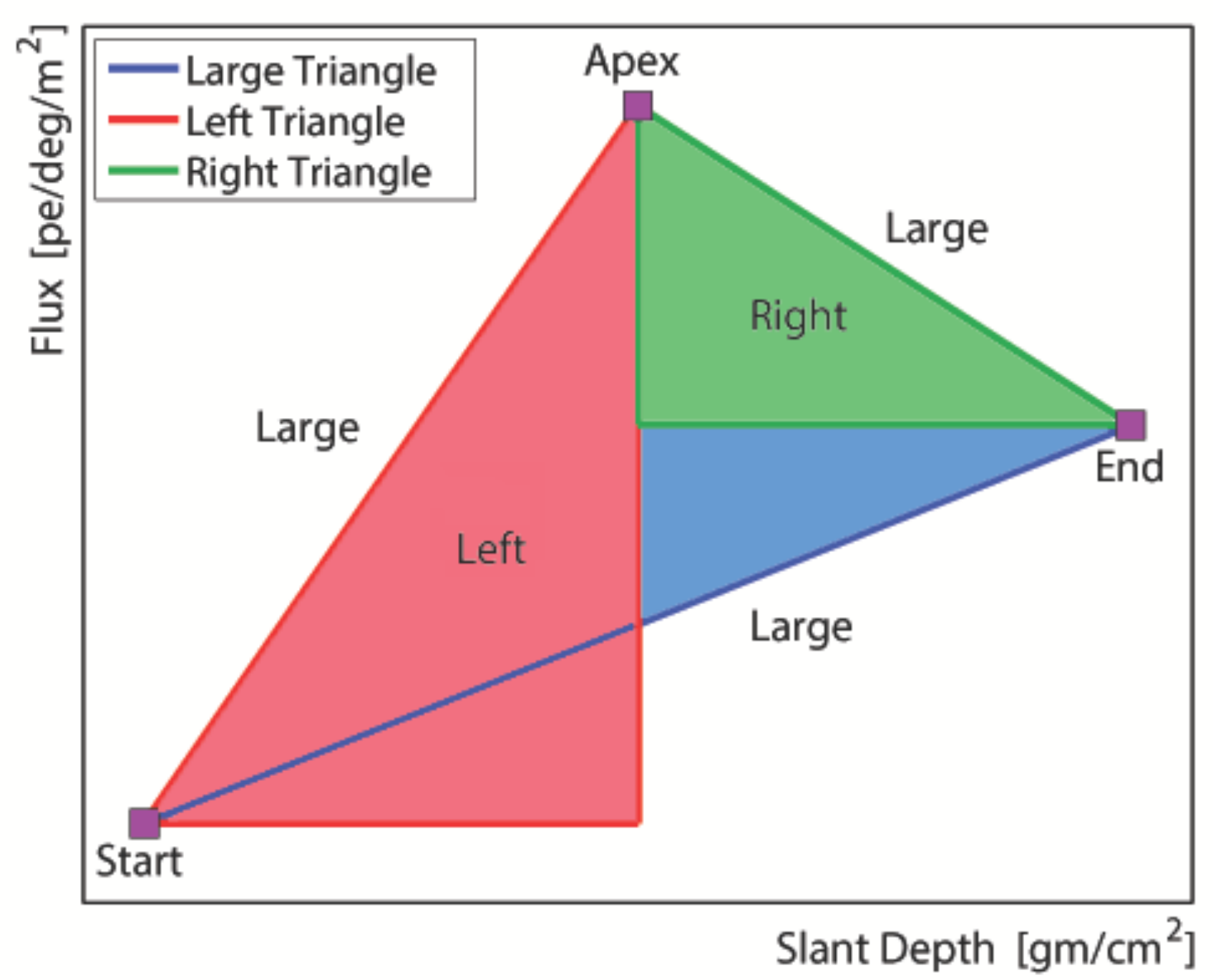}
   }
   \caption{Triangles created from the feature construction step.} 
  \label{fig:LabeledTriangle}
\end{figure}

The pattern recognition used is a non-adaptive track finder, similar to those used in particle physics analysis~\cite{RevModPhys.82.1419}. In this particular case, the ``track" is the extensive air-shower profile, and the usual detector ``track model," is the GH~function~\cite{gaisserhillas}. We use the simplest possible simplification of the GH~distribution: a triangle. All the parameters necessary to discriminate against flat events are contained in a set of triangles found from each event's binned photon flux signal versus reconstructed atmospheric depth pattern (see Figure \ref{fig:LabeledTriangle}). The pattern recognition finds, and sets limits on, the allowed shapes of the extracted triangles, and rejects events outside those limits. Only events which contain useful information remain after cuts, based on these limits, are applied.

Pattern recognition is done in two steps: training and application. Training involves training set construction, feature construction, decision tree population, and feature selection~\cite{guyon2006feature}. 

\begin{figure}[tbp!]
  \centerline{
   \includegraphics[width=0.9\linewidth]{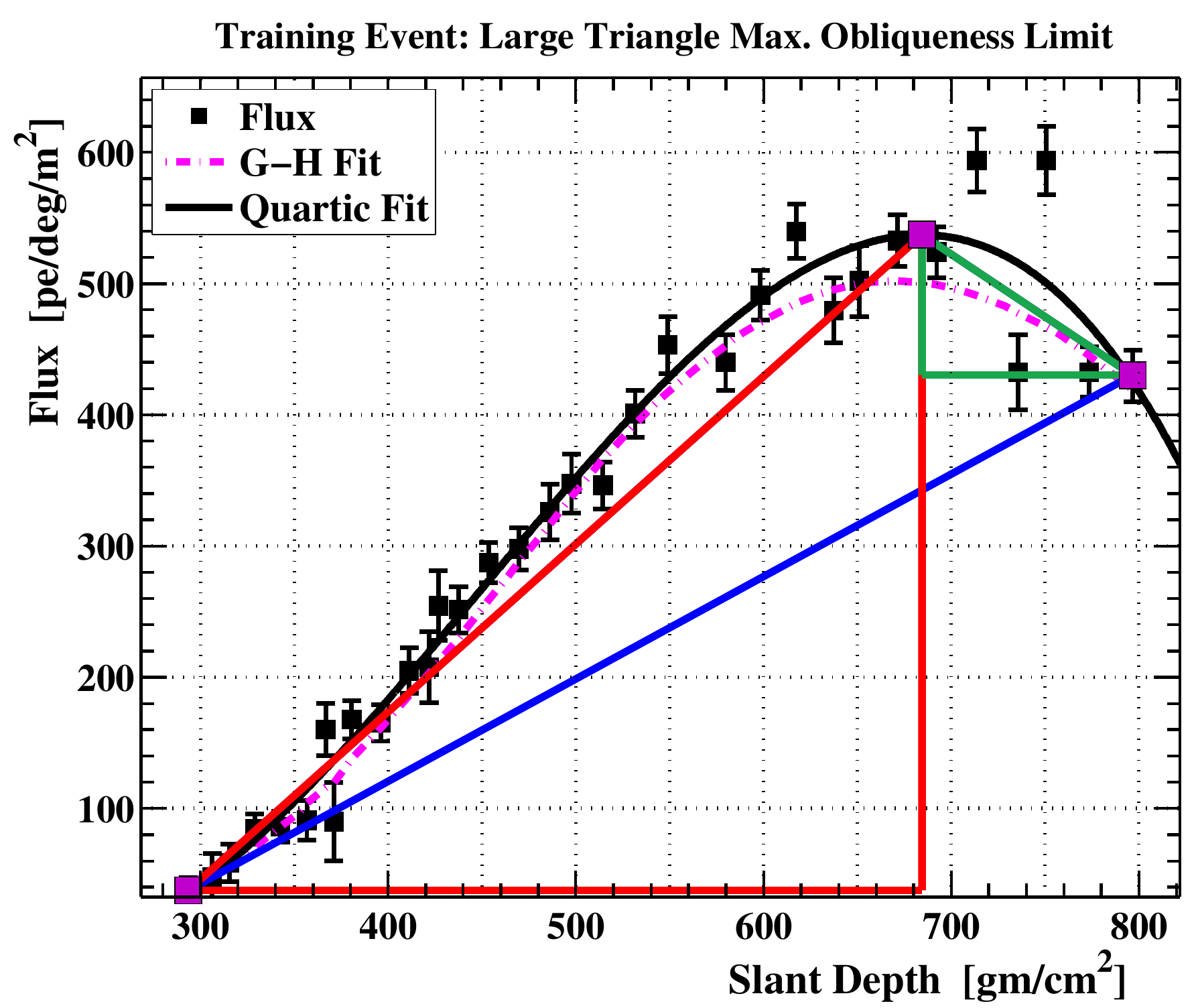}
   }
 \caption{The decision tree branch, which sets the maximum limit on the obliqueness of the large triangle, is populated by the obliqueness calculated from this event. Bins with large errors have been removed for display purposes.} 
  \label{fig:Event12}
\end{figure}

\begin{figure}[tbp!]
  \centerline{
   \includegraphics[width=0.9\linewidth]{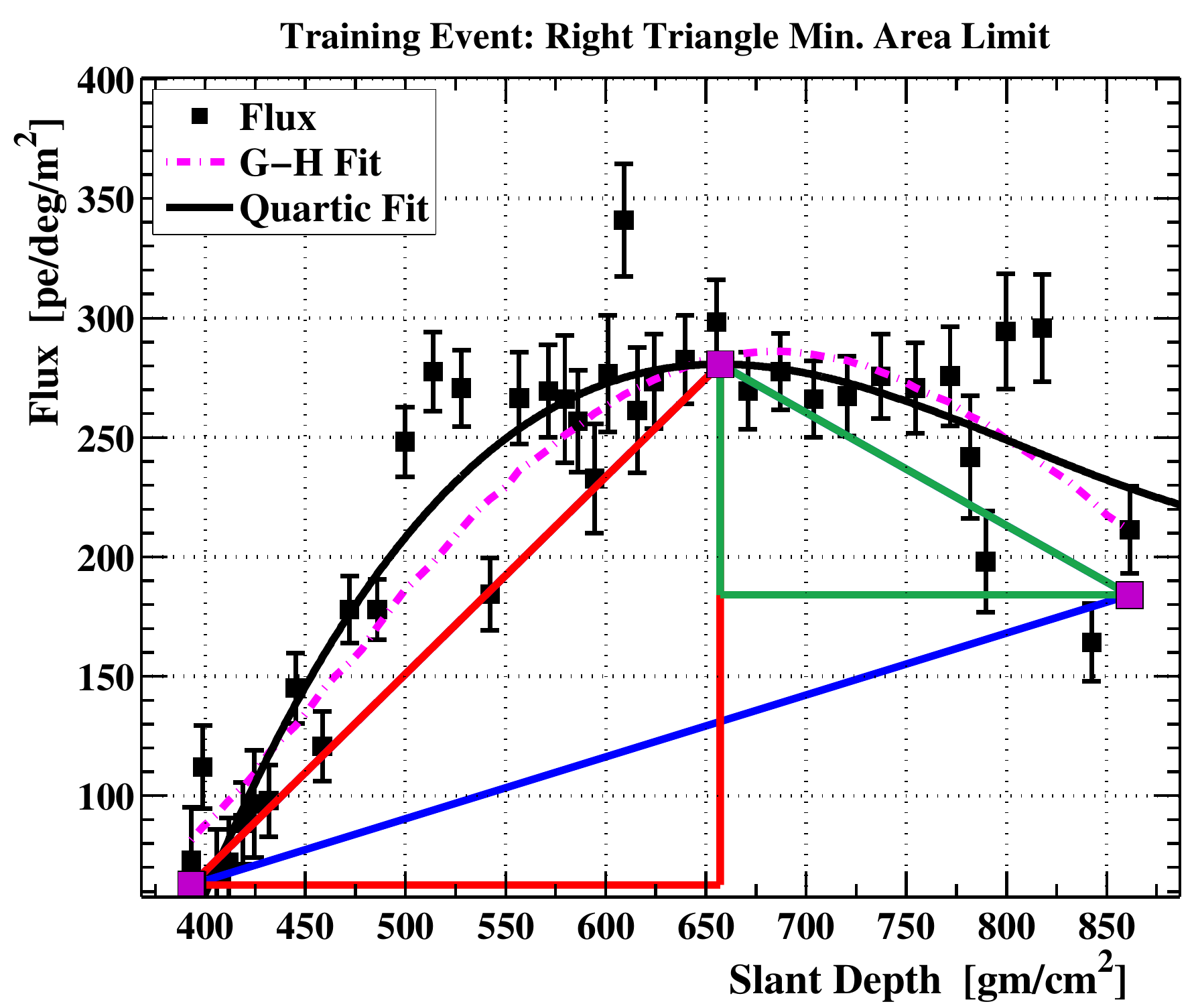}
   }
   \caption{The decision tree branch, which sets the minimum limit on the area of the right triangle, is populated by the right triangle area calculated from this event. Bins with large errors have been removed for display purposes.} 
  \label{fig:Event88}
\end{figure}

\begin{figure}[tbp!]
  \centerline{
   \includegraphics[width=0.9\linewidth]{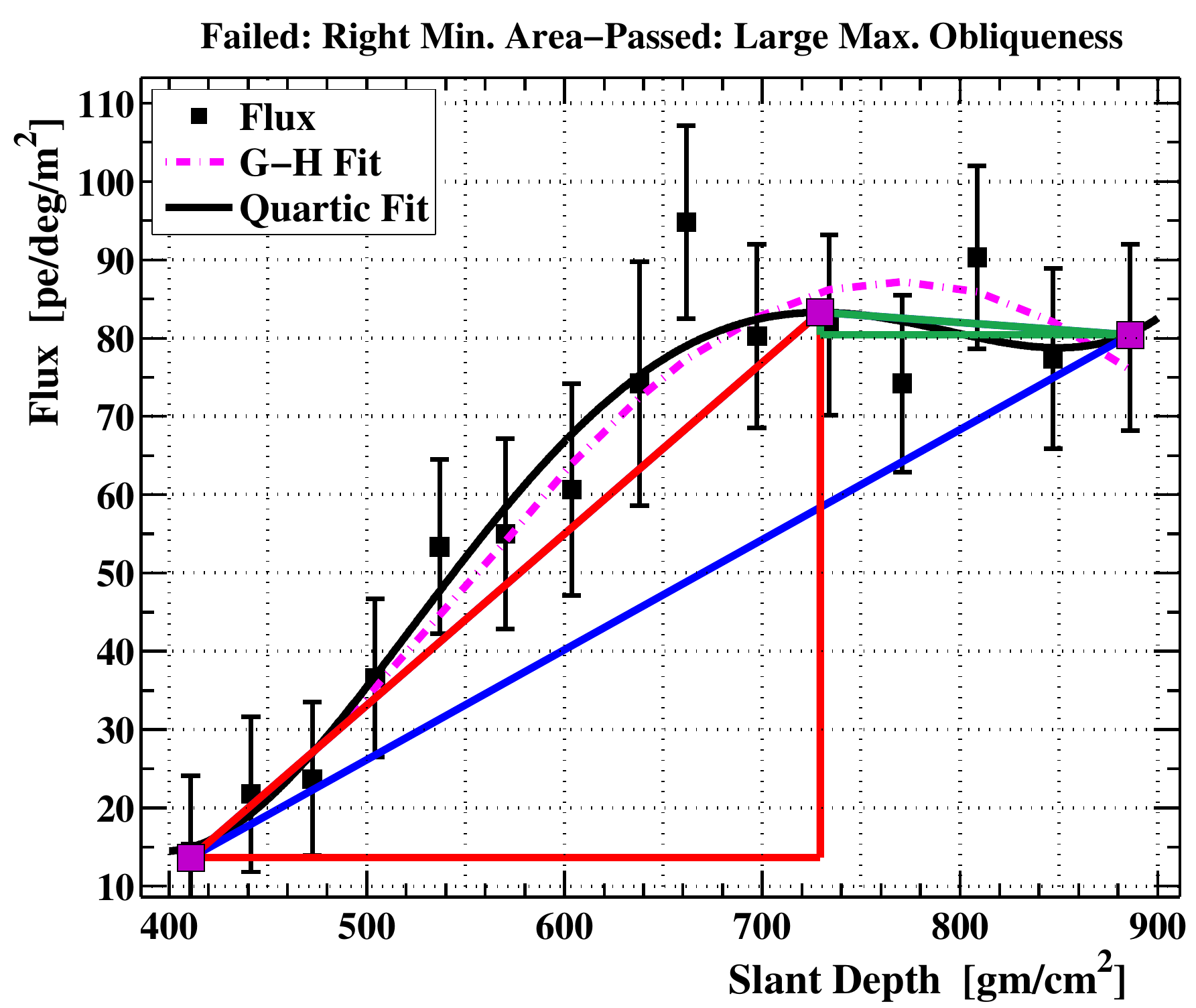}
   }
   \caption{An event which passed the large triangle obliqueness test, but failed the right triangle area test.} 
  \label{fig:Event77}
\end{figure}

\begin{figure}[tbp!]
  \centerline{
   \includegraphics[width=0.9\linewidth]{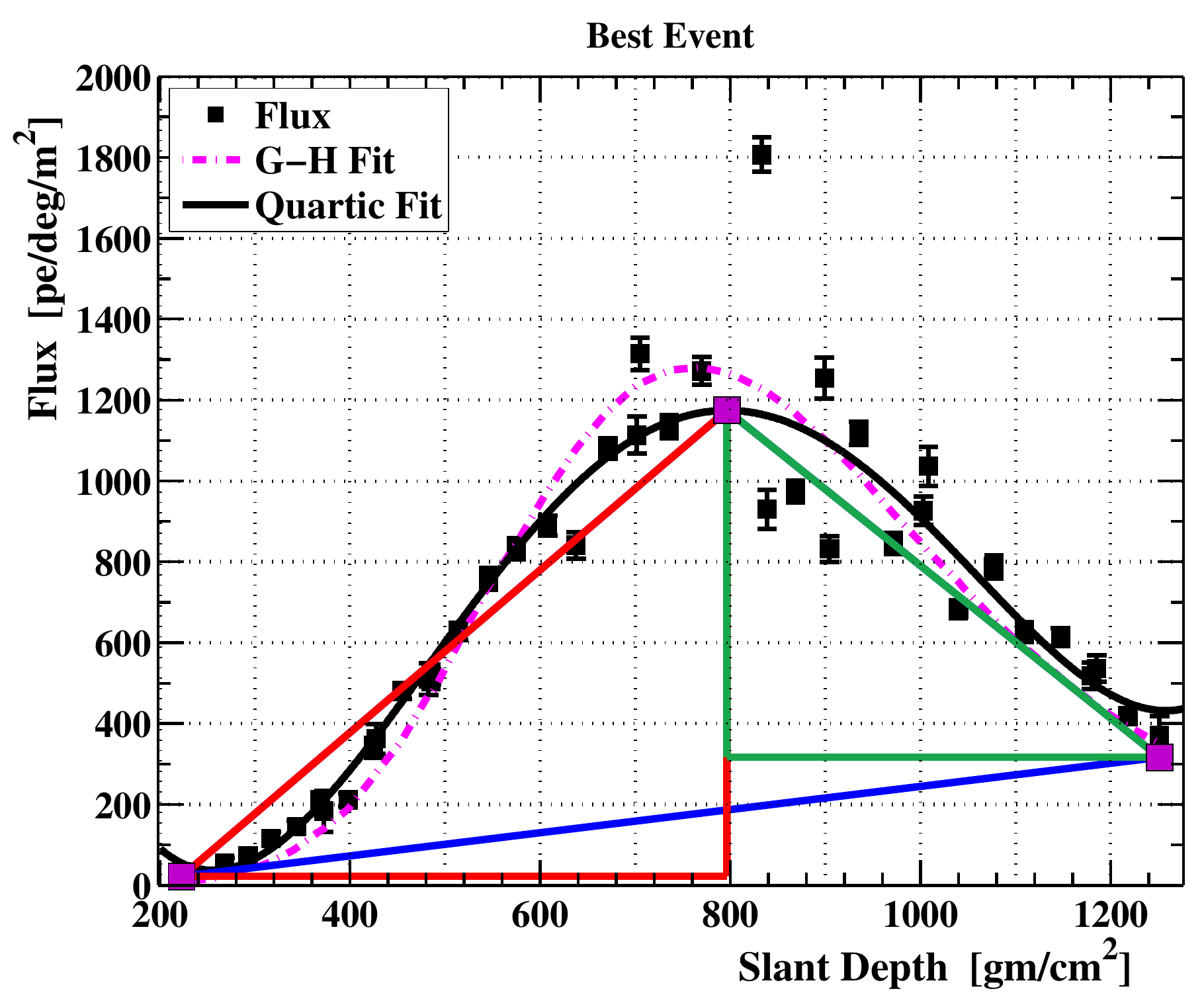}
   }
   \caption{The single passed event which has the minimum value of large triangle obliqueness, and the maximum value of the right triangle area, of the whole set. This is also the highest energy event, at $\log_{10}(\rm{Energy/eV}) = 20.12$. Bins with large errors have been removed for display purposes.} 
  \label{fig:Event1}
\end{figure}

\begin{enumerate}[(a)]
\item
Training set construction involves selecting a subset of the data, and Monte Carlo simulated, events. This set of events was scanned by eye and categorized as good based on whether a rise, and fall, can be discerned. Data, and MC simulation, are weighed equally at this point, as all events are simply used to find limits, on the triangle geometries, which can contain useful information. 

\item
Feature construction is done by finding the start, apex, and end points of each reconstructed event signal histogram in the training set.  These three points are used to form five triangles. Large, left, right, under left, and under right. The three most useful of these are labeled in Figure~\ref{fig:LabeledTriangle}. (See also Figures~\ref{fig:Event12}~to~\ref{fig:Event1}.)

\indent To find the triangle apex in the noisy signal, a fit to a quartic polynomial is done. The method used is a bisquare weights fit, which is an iteratively reweighted least-squares (IRLS), using Tukey's biweight function, that minimizes the effect of outlying bins~\cite{holland1977robust}. The apex is the local maximum of this fit, within the track of the shower.  

\indent Two other triangle vertex points are found from the shower reconstructed start/end depths and signal, calculated by a linear fit using the bisquare weights method. A linear fit is done for the start/end vertex points, because the quartic fit is unstable there, as there are more data points around the apex. This fit is done on a selection of bins at the start of the shower, before the apex, and at the end after the apex for the end point. If there are less than three bins on a side of the apex then a simple weighted average in signal and slant depth is done using the first, or last three bins, for that vertex point.

\item
The decision tree (Figure \ref{fig:DecisionTree}) is populated by calculating, for each event, a number of variables which depend on: the extracted triangles, the signal itself (mean, standard deviation, etc.), and the parameters of the signal quartic polynomial fit which was used to find the apex. The extremums of the training subset, for each of these parameter distributions, establishes the value of a branch of the decision tree. These branch limits are used as a Boolean check, for each event, of the whole data set. For example (See Figure \ref{fig:Event88}), the minimum allowed area of the right triangle is determined by the extremum value of that variable for the training set. If the value of an event is less than this it is rejected. For an event to pass, it must pass the test for all variables (See Figure \ref{fig:TwoDimPlot}).

\begin{figure}[tbp!]
  \centerline{
   \includegraphics[width=0.9\linewidth]{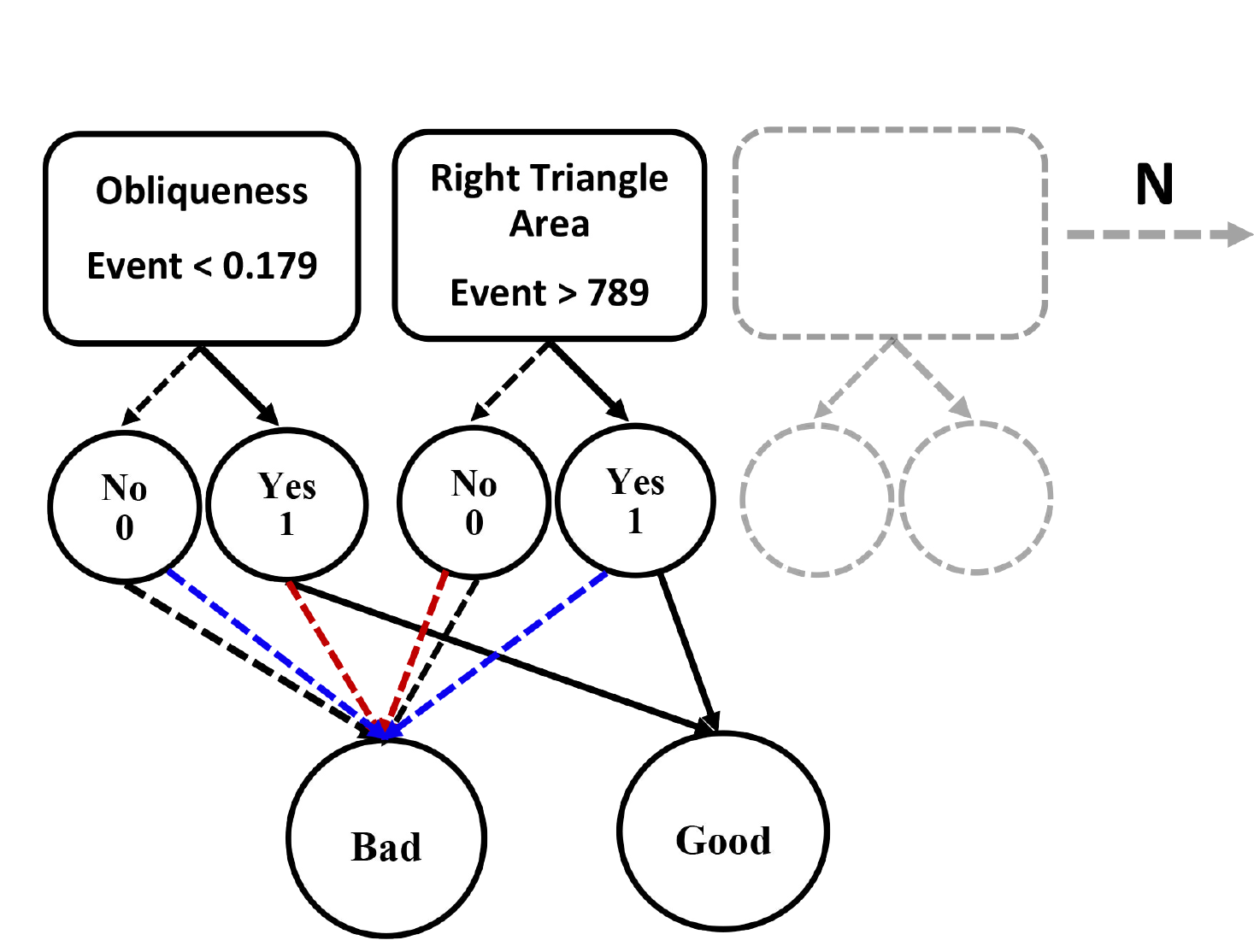}
   }
   \caption{Decision Tree Pictogram} 
  \label{fig:DecisionTree}
\end{figure}

\begin{figure}[tbp]
  \centerline{
   \includegraphics[width=0.9\linewidth]{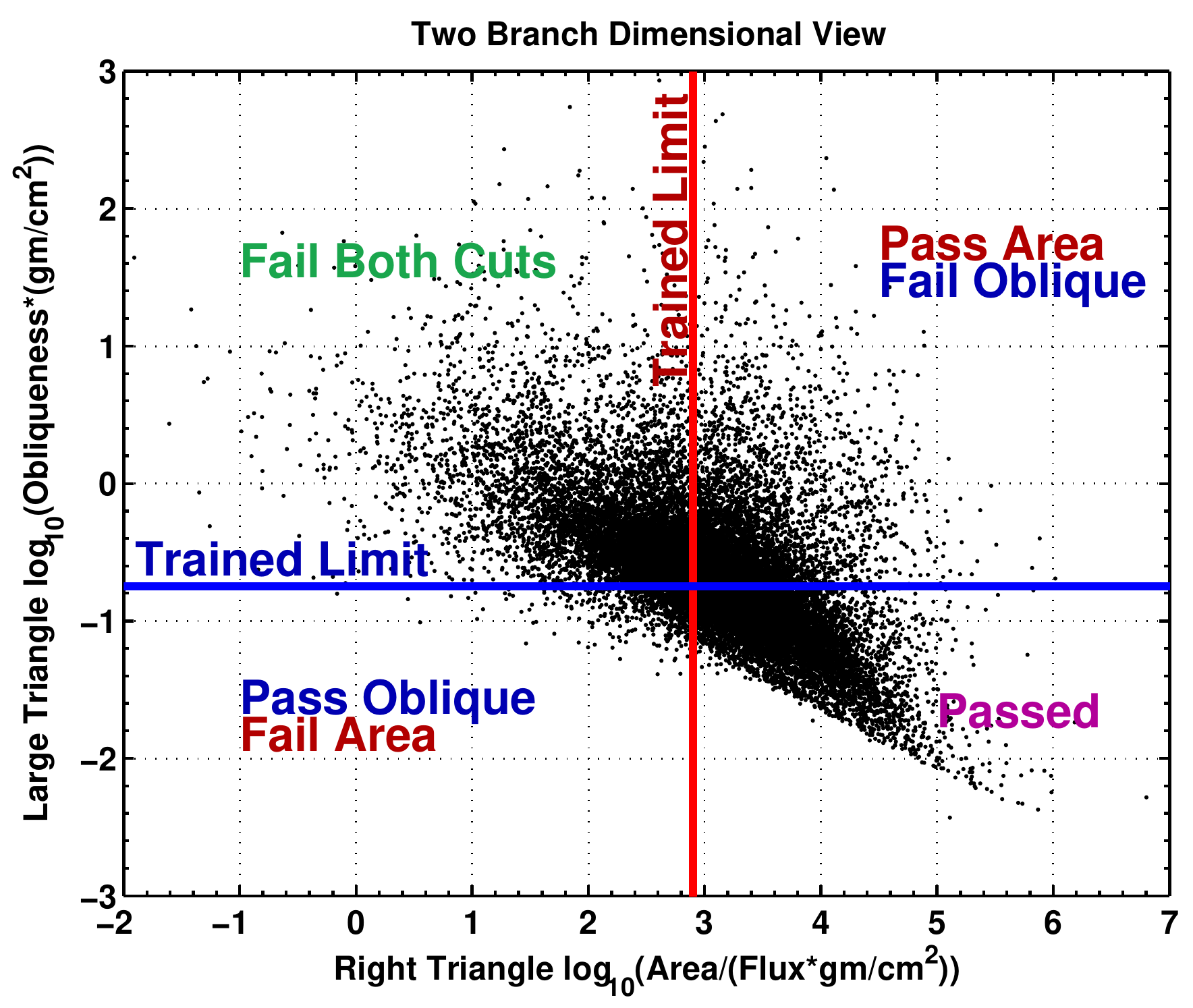}
   }
   \caption{Dimensional view of two branches. Units given should not be interpreted as physical quantities.} 
  \label{fig:TwoDimPlot}
\end{figure}

\item
Feature selection is a process which reports which of the calculated measures are sufficient to categorize events.  To minimize the number of calculated parameters needed to categorize events as good, or bad, and reduce the number of false negatives (resulting from overfitting), when applied to the whole data set, variables that remove less than .5\% of the training set, and single variables (or groups of variables) that remove the same events as another variable (or group of variables), are pruned from the decision tree.  The full method will be explained in detail in \cite{Lundquist2014}.

\end{enumerate}

\begin{figure}[tbp!]
  \centerline{
   \includegraphics[width=0.9\linewidth]{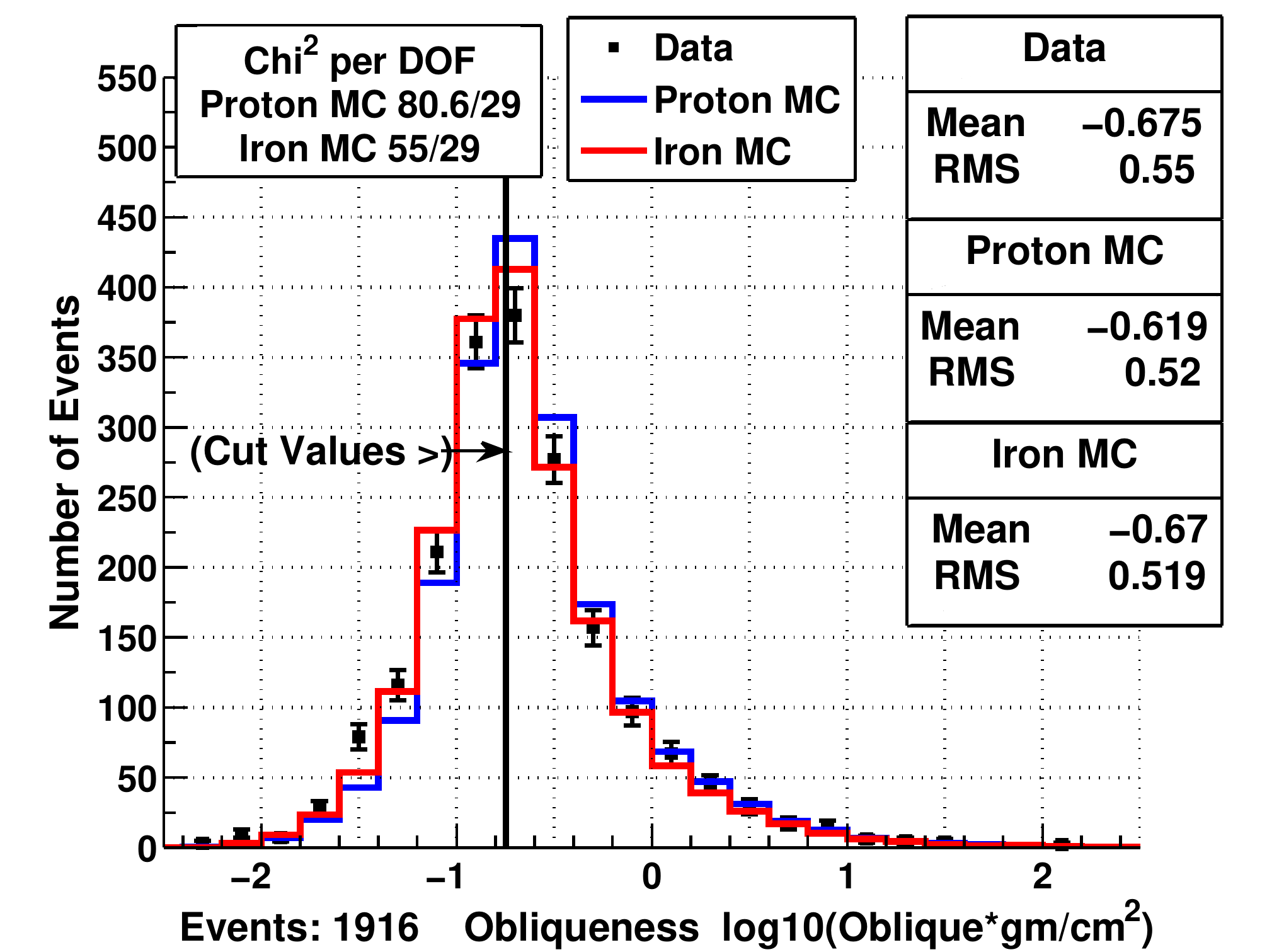}
   }
   \caption{Data/Proton MC comparison, for all events with no cuts, of the obliqueness of the large triangle. The chi-squared test value shows agreement between QGSJETII-03 MC and data. Units given should not be interpreted as physical quantities.} 
  \label{fig:AllObliqueDist}
\end{figure}

\begin{figure}[tbp!]
  \centerline{
   \includegraphics[width=0.9\linewidth]{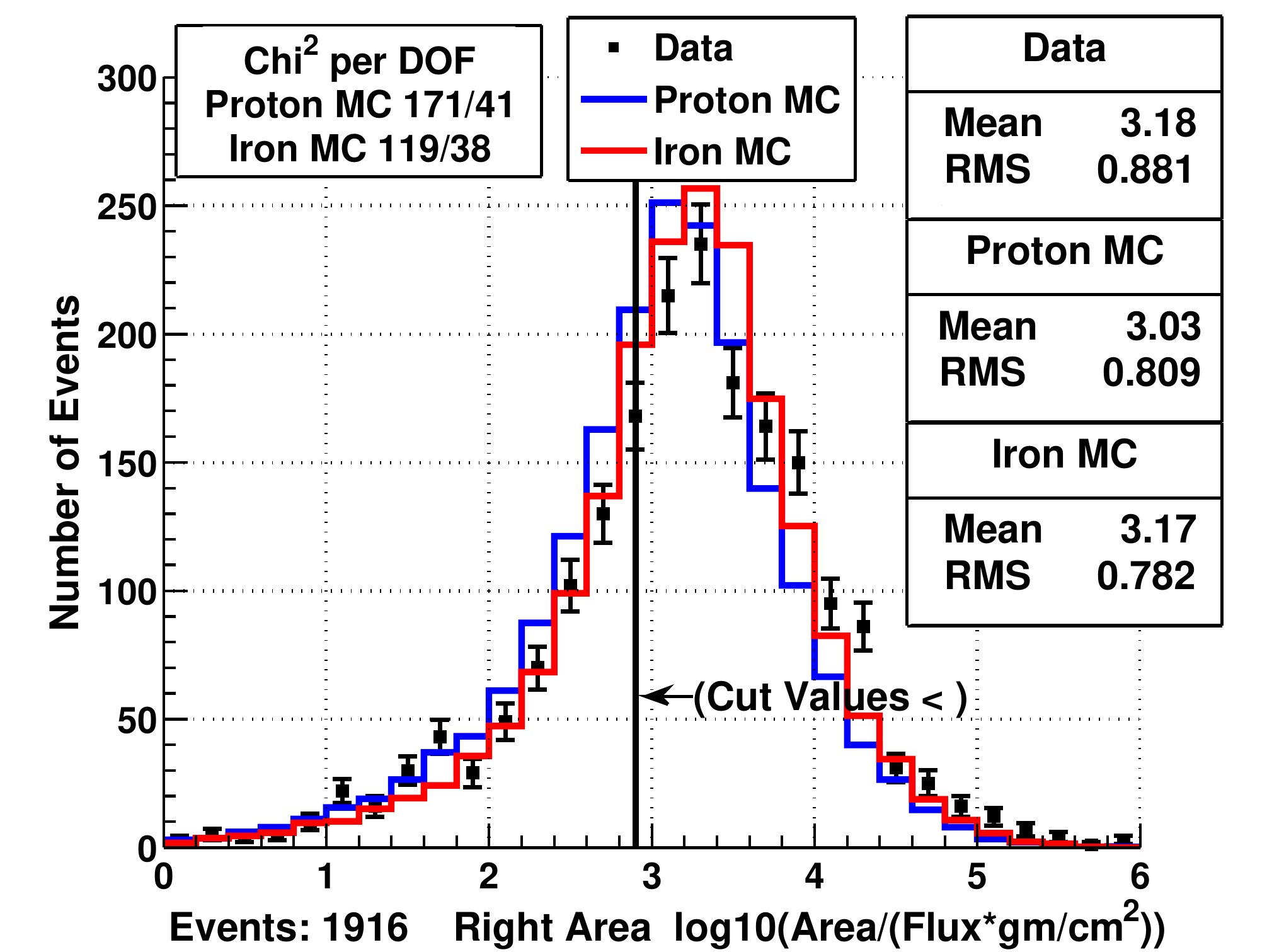}
   }
   \caption{Data/Proton MC comparison, for all events with no cuts, of the area of the right triangle. The chi-squared test value shows agreement between QGSJETII-03 MC and data. Units given should not be interpreted as physical quantities.} 
  \label{fig:AllRightArea}
\end{figure}

The two most effective cuts, those that remove the highest number of bad events when applied individually, are: a maximum limit to the allowed perimeter/area (this is called the obliqueness) of the large triangle, and the minimum allowed area of the right triangle (see Figure~\ref{fig:LabeledTriangle}). Distributions of these parameters for all events, with no cuts applied, are shown in Figures~\ref {fig:AllObliqueDist} and~\ref{fig:AllRightArea}. The binned maximum likelihood estimated chi-squared test values~\cite{Baker1984}, for the proton/data comparisons are shown on each plot.  The agreement between data and proton MC validates the use of combining both sets for pattern recognition training. Figures~\ref{fig:Event12}-\ref{fig:Event1} show examples of events and their branch characteristics. 

For the training subset of eye scanned good events, the event in Figure~\ref{fig:Event12} was found to have the maximum value of the obliqueness of the large triangle. The obliqueness of this event populates the branch which sets the limit on the maximum allowed obliqueness of all events. The training subset good event in Figure~\ref{fig:Event88} was found have the minimum value of the area of the right triangle. The area of the right triangle of this event populates the branch which sets the limit on the minimum allowed right triangle area of all events. Figure~\ref{fig:Event77} shows a failed event, for which the obliqueness is less than the maximum allowed obliqueness, but the right triangle area is smaller than the allowed limit. Figure~\ref{fig:Event241} shows a failed event, for which the right triangle area is greater than the minimum allowed right triangle area, but the large triangle obliqueness is larger than the allowed limit. Figure \ref{fig:Event1} shows a good event which passes both cuts.

Applying the resulting decision tree to the training set, results in zero false negative events (found bad, when eye scan said they were good). This is by definition, as the tree is populated by the limits of each variable found from the training set. Since the categorization by eye may not be perfect, false positive events (found good, when eye scan said they were bad), are possible when applied to the training set. The number of false positives, on the 399~event data training subset, was 1.2\%. The number of false positives, on the 412~event proton MC training subset, was 2.9\%. This results in an overall accuracy of 97.6\% on the training set.

Application involves extracting the features for the set of all data and proton/iron simulated events, calculating the parameters which survived the feature selection process, and applying the decision tree. If the particular parameter calculated for each event is within the required limit of the branch that makes a decision on that variable, that particular branch passes that event. If an event is passed by all branches, it is considered a good event.

The result is a set of events with peaks far enough away from both edges, and a sufficient amount of curvature of the signal from the peak to either edge, so that we can be confident that \xmax\ is within the field of view.

Random test samples of~200~events were selected from the data and proton MC sets. The result was that the pattern recognition is 96.5\% accurate on the data test set, with 3~false positives, and 4~false negatives. The accuracy on the proton MC test set is also 96.5\%, with 2~false positives, and 5~false negatives. Twice as many random events~(400) were chosen for the iron MC test set, as the pattern recognition was not trained on iron MC events. The iron MC accuracy is also 96.5\%, with 1~false positive, and 13~false negatives.

The overall accuracy when comparing the eye scan and pattern recognition, for both training and test sets, is 97.2\%, when including false positives and negatives. Since events in which \xmax\ is not seen are on average not reconstructed well, and the improvement in resolution (See Figure \ref{fig:resolutions}) shows us that, on average, events in which there is clear \xmax\ in view are well reconstructed, only false positives are on average detrimental to the \xmax\ resolution. If pattern recognition is considered inaccurate only for false positives the accuracy percentage becomes~99.6\%. 

\section{Geometrical Cuts Optimized for Pattern Recognition Events and their Effect on Resolution}

The pattern recognition cuts were applied on all data that passed the weather cut, and all MC events, with no prior geometrical cuts. However, due to the fact that incorrectly reconstructed events can still have a clear \xmax\ in view, and to further improve \xmax\ and energy resolution, cuts which take into account the geometry of the events need to be applied, in addition to the pattern recognition cut. These geometry cuts were optimized using the proton MC resolutions and biases, with the priority being minimizing the energy dependence of the \xmax\ resolution. 

To find the optimal set of cuts several considerations were taken into account: These include improving the overall \xmax\ resolution, minimizing the change of \xmax\ resolution with energy, improving the overall energy resolution, and minimizing the dependence of energy resolution on energy, while maximizing the total number of events. Secondary to these considerations were minimizing the \xmax\, and energy biases, produced by applying the cuts. 

The resulting final set of optimized geometry cuts, applied to the events that passed the pattern recognition cuts, are listed below. Events which satisfy these inequalities are removed from the data set.

\begin{enumerate}
\item Zenith angle $>$ 58$\Deg$
\item Boundary Distance $<-$500~m (negative values are outside the array)
\item Hybrid/Surface Core Difference $>$ 1600~m
\item Geometry Fit $\chi^2$/DOF $>$ 4.5
\item Start \xmax\ Bracket: (\xmax\ - $X_{start}$) $<$ 20~\gcm
\item End \xmax\ Bracket: ($X_{end}$ - \xmax) $<$ 0~\gcm
\item Energy $ < 10^{18.2}$~eV
\end{enumerate}

Note that these are similar to, but looser than, the original geometrical cuts. For instance, the MC shows that if the shower core is just outside the array, the event can still be reconstructed accurately if \xmax\ is clearly in view of the FD. Previously, events which fell outside of the array were cut.

\begin{figure}[tbp!]
\centerline{
\includegraphics[width=0.8\linewidth]{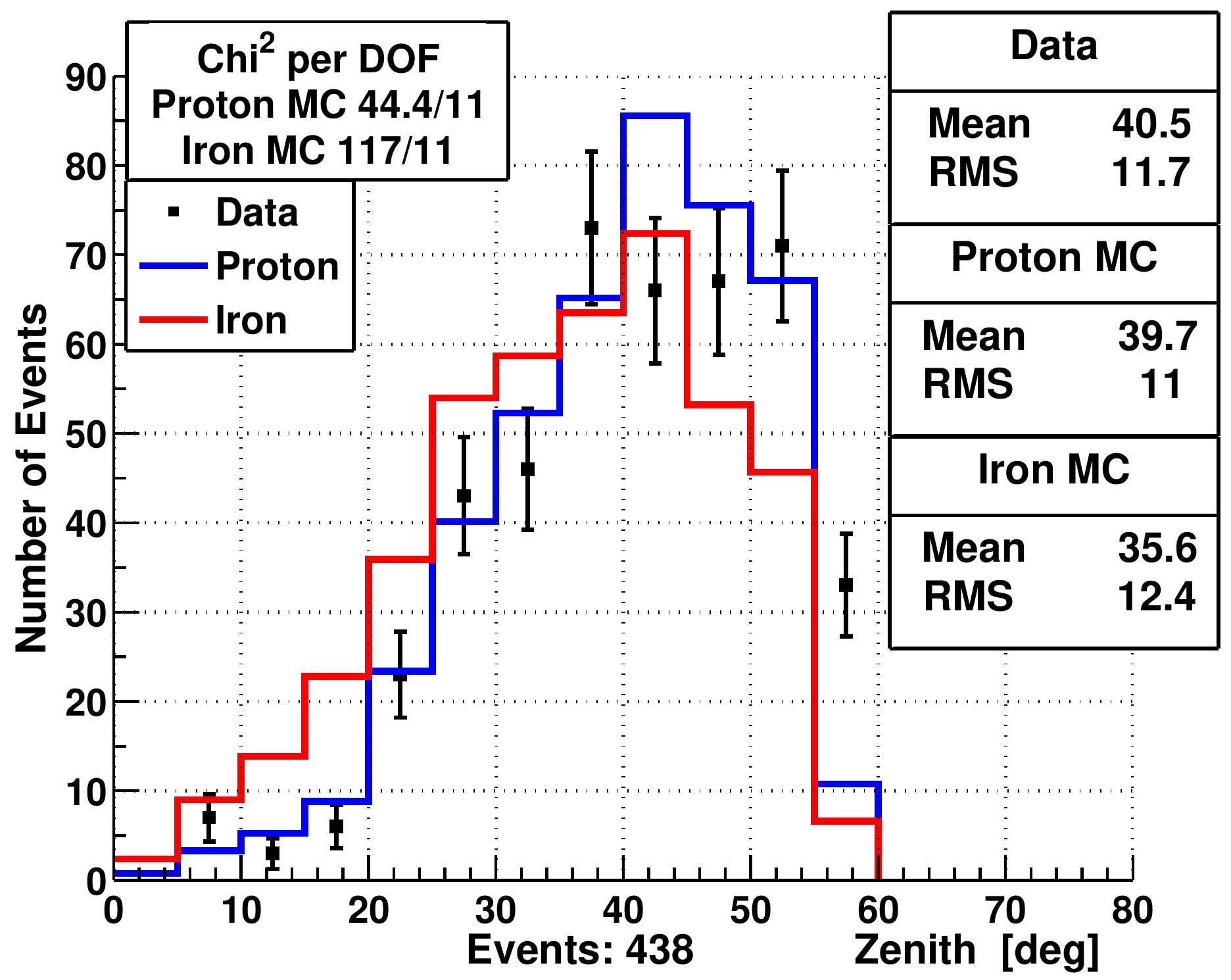}
}
\caption{Zenith angle Data/MC comparison, for events which pass the pattern recognition with geometry cuts. This distribution is expected to have a composition dependence. The chi-squared test value shows better agreement between QGSJETII-03 proton MC and data, than iron.} 
\label{fig:zenithdist}
\end{figure}

\begin{figure}[tbp!]
\centerline{
\includegraphics[width=0.8\linewidth]{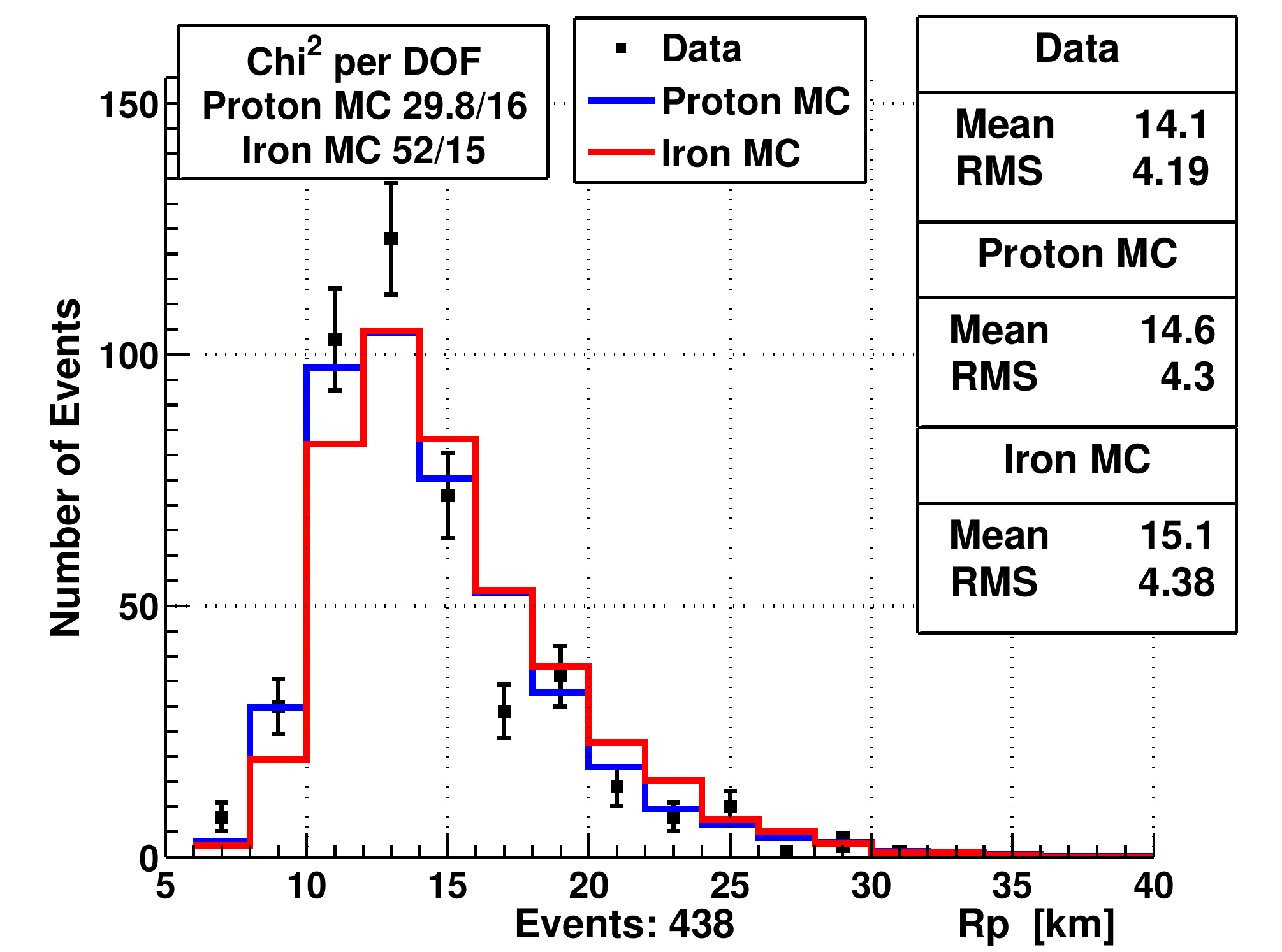}
}
\caption{$R_P$ (distance of closest approach of shower to FD) Data/MC comparison, for events which pass the pattern recognition with geometry cuts. The chi-squared test value shows agreement between QGSJETII-03 MC and data.} 
\label{fig:rpdist}
\end{figure}

\begin{figure}[tbp!]
\centerline{
\includegraphics[width=0.8\linewidth]{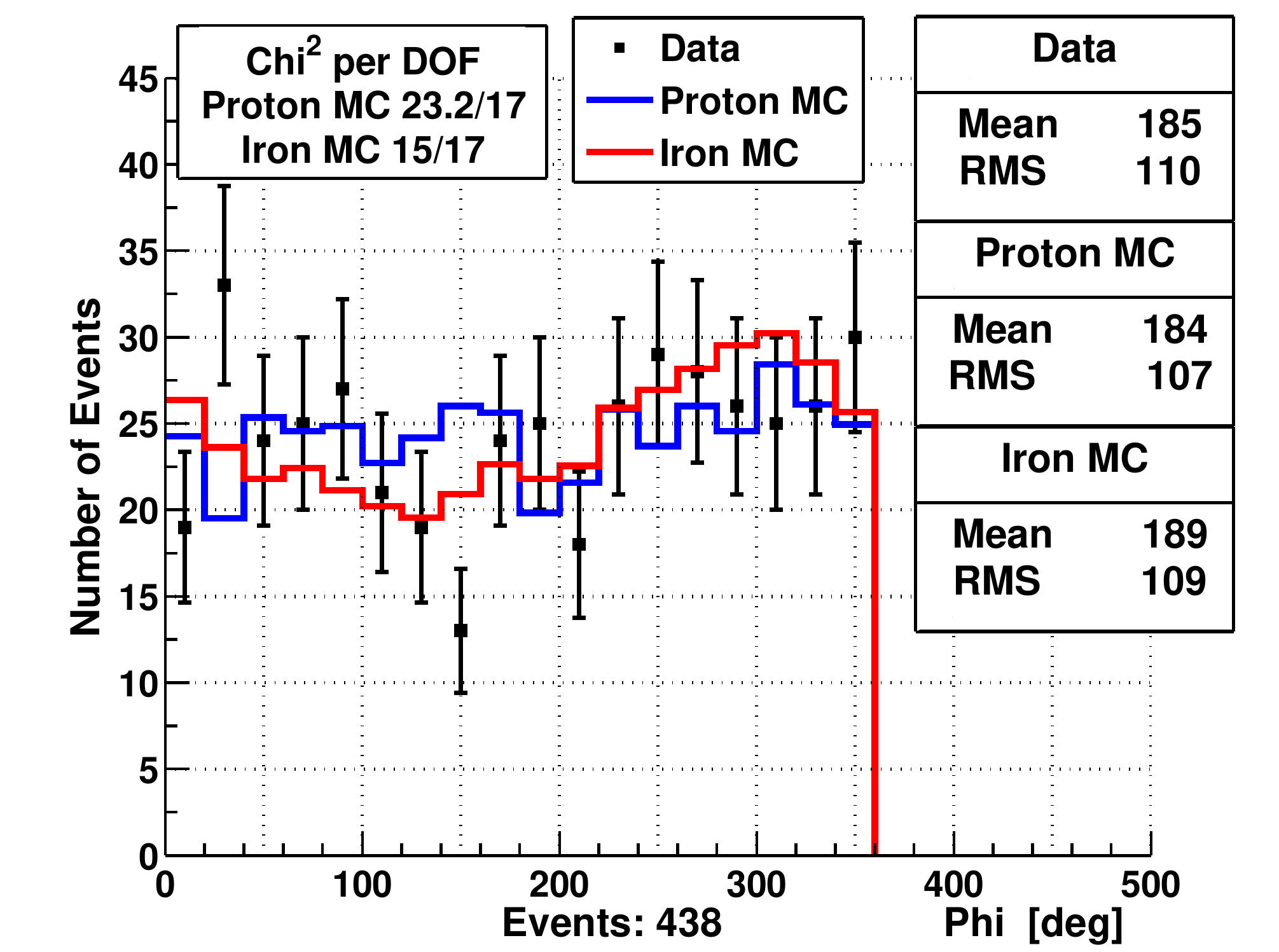}
}
\caption{Phi (azimuthal) angle Data/MC comparison, for events which pass the pattern recognition with geometry cuts. The chi-squared test value shows good agreement between QGSJETII-03 MC and data.} 
\label{fig:phidist}
\end{figure}

\begin{figure}[tbp!]
\centerline{
\includegraphics[width=0.8\linewidth]{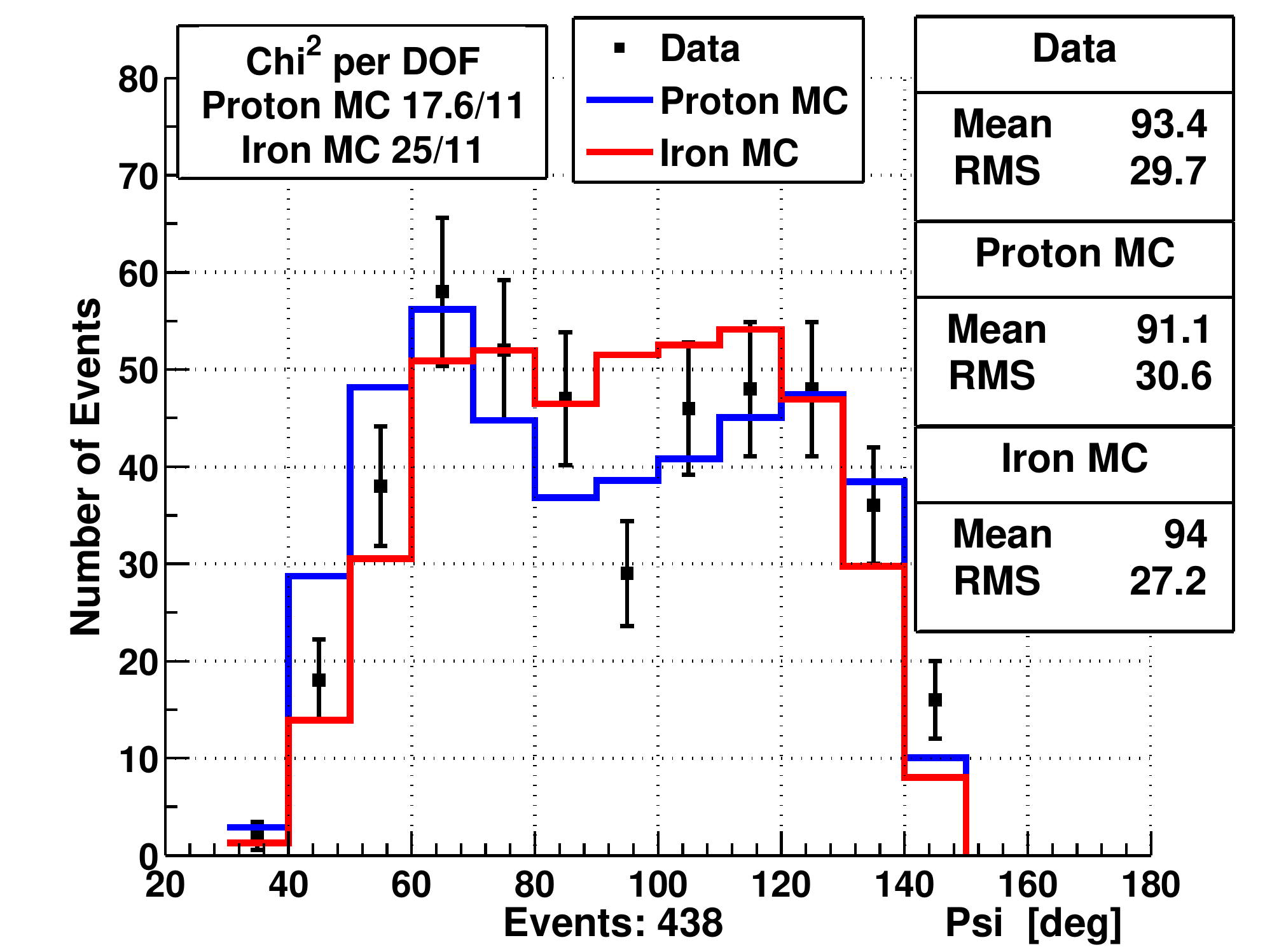}
}
\caption{Psi (shower-detector plane) angle Data/MC comparison, for events which pass the pattern recognition with geometry cuts. The chi-squared test value shows good agreement between QGSJETII-03 MC and data.} 
\label{fig:psidist}
\end{figure}

To ensure that the detector is accurately modeled, and biases are not introduced with these cuts, some comparisons between MC and data, for shower variable distributions other than \xmax, are needed. The zenith angle distribution (Figure \ref{fig:zenithdist}), which is expected to have some composition dependence, shows that there is better agreement between the data and proton MC, than between data and iron MC. This effect has been seen previously \cite{Belz:06/10dba}. Distribution comparisons of $R_P$ (distance of closest approach of shower to FD), phi (azimuthal angle), and psi (angle in shower-detector plane) can be found in Figures~\ref{fig:rpdist}-\ref{fig:psidist}. Data and MC are in good agreement for all of these distributions. Comparisons between data and MC are also in agreement as a function of energy.

\begin{figure}[tbp!]
\centering
\subfloat[Subfigure 1][]{
\includegraphics[width=0.8\linewidth]{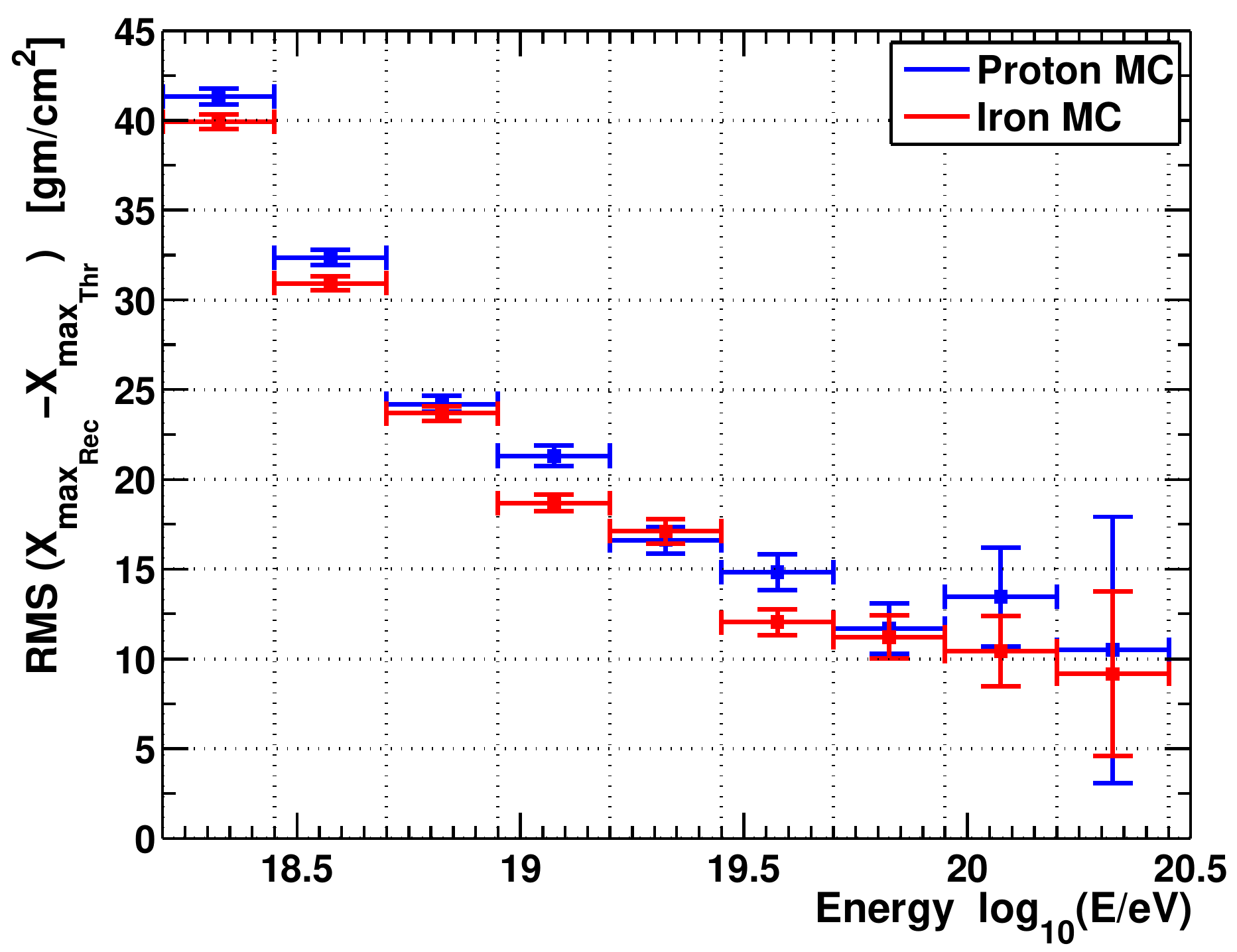}
\label{fig:monicaresolution}}
\qquad
\subfloat[Subfigure 2] [] {
\includegraphics[width=0.8\linewidth]{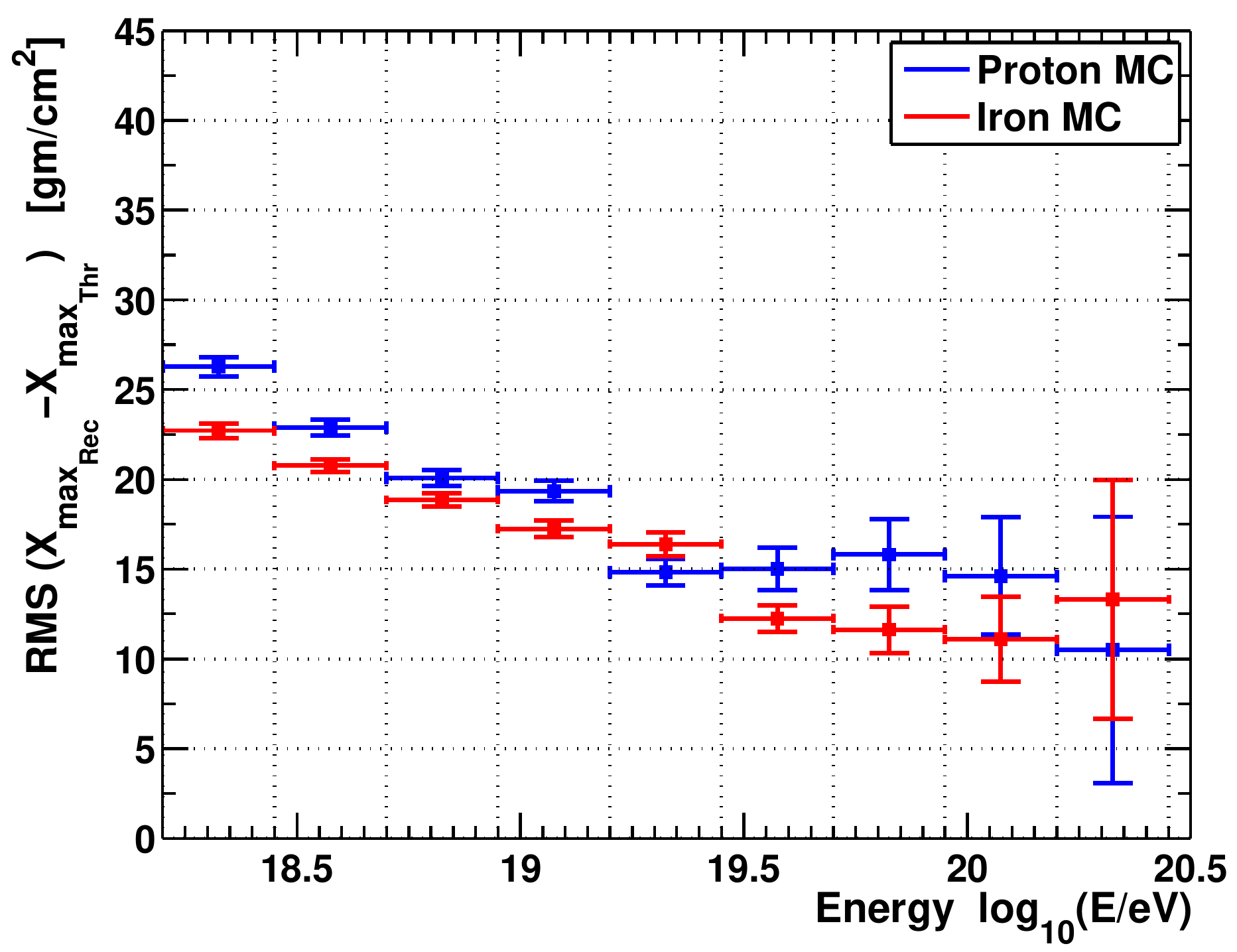}
\label{fig:optresolution}}
\caption{\xmax\ resolution plots showing the energy dependence of the RMS of the difference between QGSJETII-03 MC reconstructed and thrown \xmax. The top figure is the simple geometry cuts. Bottom figure is the pattern recognition with geometry cuts.} 
\label{fig:resolutions}
\end{figure}

\begin{figure}[tbp!]
\centerline{	
\includegraphics[width=0.8\linewidth]{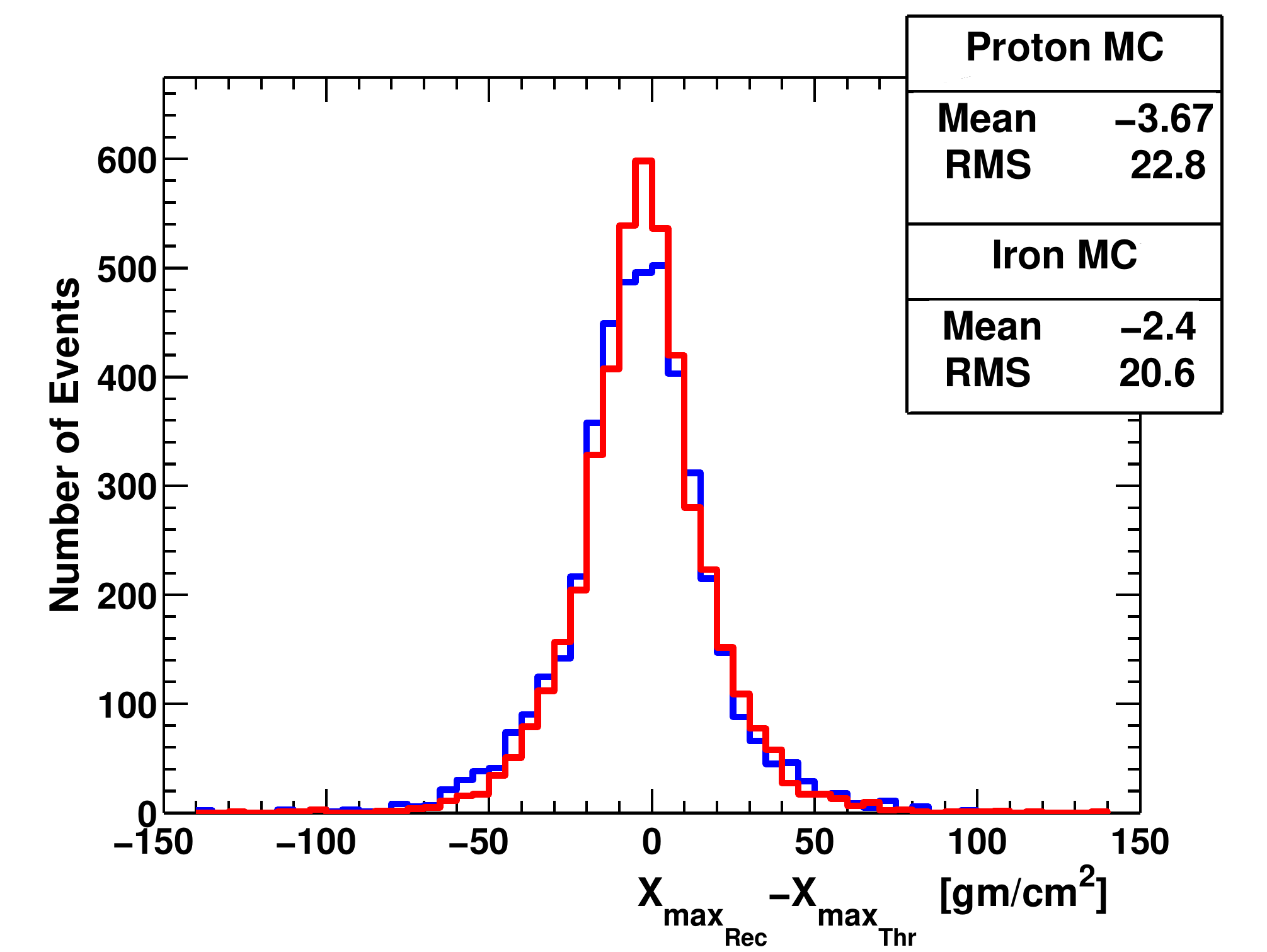}
}
\caption{The final MD hybrid \xmax\ resolution, above E~$>$10$^{18.2}$~eV, for the Monte Carlo sets: shown is the difference between the reconstructed \xmax\ values and the thrown \xmax\ values, for QGSJETII-03 MC proton induced showers (blue), and iron induced showers (red).} 
\label{fig:FinalResolution}
\end{figure}

The effect of geometry cuts being applied, in addition to the pattern recognition cuts, gain little by way of improving the overall energy and \xmax\ resolution, but are used to stabilize fluctuations at all energies, and further decrease the slope of the resolution dependence on energy. A comparison of the \xmax\ resolution dependence on energy, for the simple geometry cuts and the final result of geometry cuts with pattern recognition, is shown in Figure~\ref{fig:resolutions}. The final achieved overall resolutions are shown in Figure~\ref{fig:FinalResolution}. The resolution of the proton set is 22.8~\gcm, with a bias of $-$3.67~\gcm. Resolution of the iron set is comparable, with a width of 20.6~\gcm\, and a bias of $-$2.4~\gcm.

The total number of data events which pass all cuts (pattern recognition and geometrical) is 438, corresponding to 22.8\% of 1916~reconstructed events in the five year time period. The percentage of proton MC events that pass all cuts is~24.0\%, and the percentage of iron MC events that pass all cuts is 27.8\%. Though there are fewer events, compared to the simple geometry cuts alone, there are an equal number of events with energy~$>$10$^{19.2}$~eV.

\begin{figure}[p!]
  \centerline{
   \includegraphics[width=0.9\linewidth]{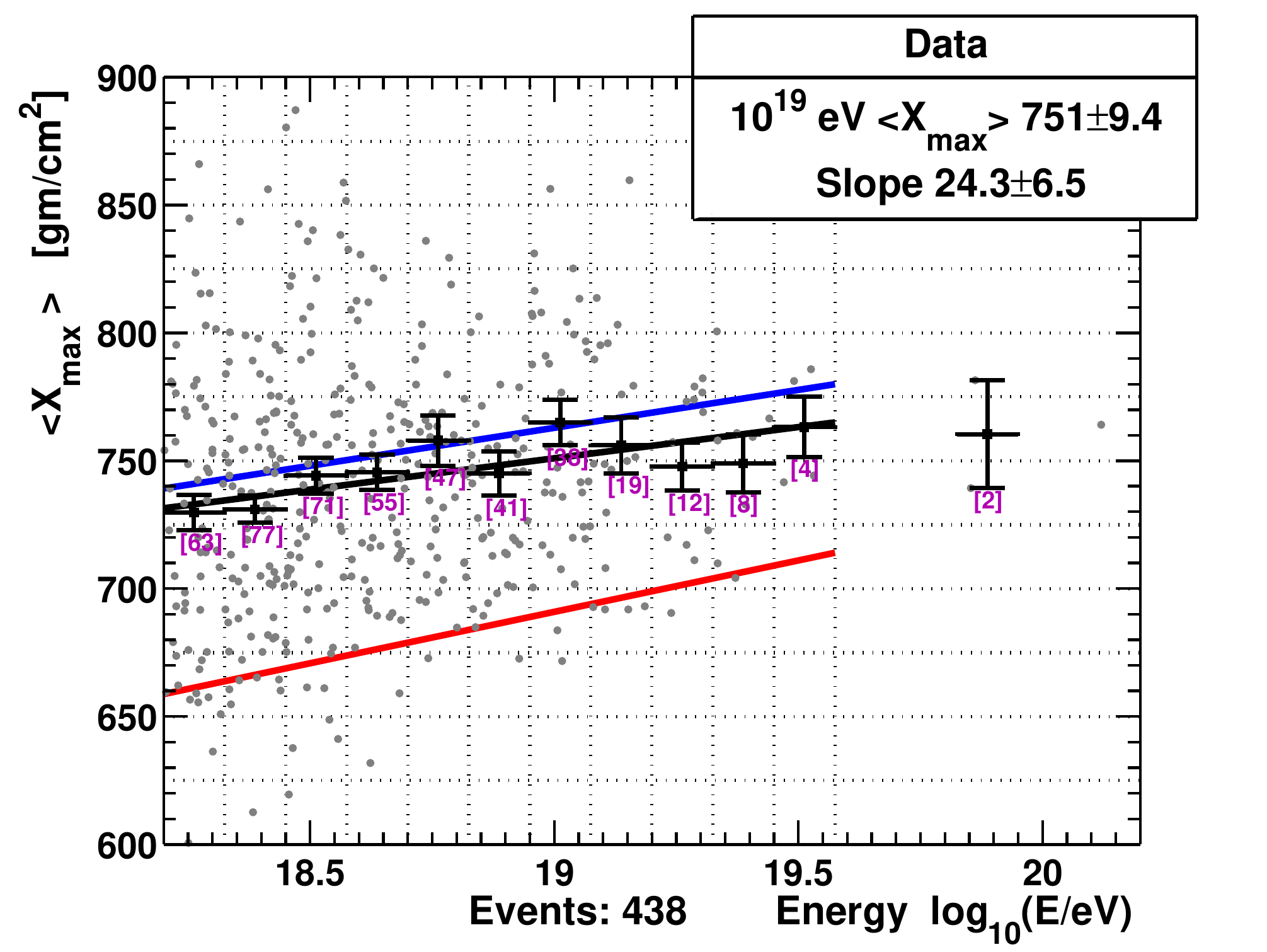}
   }
   \caption[The final Middle Drum hybrid composition result using geometry and pattern recognition cuts: the \xmax\ values for each data event are plotted as a function of energy overlaid with the proton and iron ``rails". ]{The final five year Middle Drum hybrid composition result, using pattern recognition and geometry cuts: the \xmax\ values (grey points) of each data event, are plotted as a function of energy, overlaid are QGSJETII-03 proton (blue), and iron (red) MC ``rails". Black data points with error bars represent the data $<$\xmax$>$ values, in 12~energy bins (of width $\log_{10} (E/eV) = 0.125$), that are plotted as a function of bin energy. The black rail is a fit to these binned values. All rails are fitted up to the energy for which the data has low statistics. The scale is chosen for emphasis on elongation, this cuts 6 events from the scatter plot.} 
  \label{fig:ERateTotalTri}
\end{figure}

\section{Discussion of Biases and Systematics}\label{sec:systematics}
Systematic errors in geometrical reconstruction, particularly biases in zenith angle, will result in shifts in the mean \xmax. We study these effects using simulated Monte Carlo data. The overall \xmax\ resolution of proton is 22.8~\gcm, and for iron is 20.6~\gcm. There is a mean shift in zenith angle, for simulated events passing pattern recognition with geometrical cuts, which is $-$0.02$\Deg$ for protons, and $-$0.21$\Deg$ for iron nuclei. The shift in \xmax\ reconstruction, over the entire energy range, is $-$3.67~\gcm\ for protons, and $-$2.4~\gcm\ for iron. The slope per energy decade of this shift is $-$7.26~\gcm\ for protons, and $-$7.23~\gcm\ for iron. 

Direct, and scattered, Cherenkov light must be subtracted to determine \xmax. The systematic effects of uncertainties in this subtraction have been previously studied, and have been found to be negligible~\cite{Abbasi2005}. This is even more the case for this data set, since the hybrid trigger makes it difficult for the fluorescent detector to see the shower at small angles to the shower axis, minimizing the effect of direct Cherenkov light subtraction.

Another source of systematic error, that is not addressable in MC simulation, is mirror alignment. Mirror surveys have been done using star alignments, and compared to theodolite based measurements. We estimate that mirror directions are known to $\pm$0.05$\Deg$. For an average shower, at average core distance, and mean zenith angle, the resultant uncertainty in mean \xmax\ is 2.6~\gcm. This ranges from 1.0 to 3.0~\gcm over the energy range of the data.

The impact of uncertainties in atmospheric density profile has been estimated, by comparing the atmosphere used in this analysis (a yearly average over TA using radiosonde data), with the United States 1976 Standard Atmosphere (used as a reference standard in Fly's Eye, HiRes and PAO experiments). This contributes an additional uncertainty on the mean \xmax\ of 11.7~\gcm. The systematic error, on the elongation rate, due to the uncertainty of the atmospheric density profile is 3.8~\gcm~per energy decade. A complementary study of the systematic error in \xmax\ determination due to temporal, and spatial, variations in atmospheric conditions using radiosonde data from the Salt Lake City airport yields a similar error of less than 10~\gcm~\cite{Machida2005} 

The contribution of aerosols to atmospheric extinction, between the EAS and the FD, is assumed to be constant in time in this analysis, and corresponds to a vertical aerosol optical depth (VAOD) of~0.04. This is the same value that was used in the HiRes analysis. Lidar data taken at the TA site over the course of the data collection confirms this average value within experimental errors. The effect of night-to-night variations of the VAOD, as determined by Lidar data, has been previously studied~\cite{Archbold2002}. For the clear night weather cut it is estimated to contribute a systematic error of 2.0~\gcm\ on $<$\xmax$>$, and a net shift in the elongation rate of 2.2~\gcm /decade.

The total average systematic error on mean \xmax\ that is not accounted for by the reconstruction is 16.3~\gcm. This varies from 15.1 to 18 \gcm over the range of energies of the data. The total systematic error, not accounted for by the reconstruction, on the elongation rate is 3.8~\gcm~per energy decade.

We estimate that the contribution to the \xmax\ resolution due to nightly variation of aerosols is 3~\gcm\, while the seasonal variation of the atmospheric density profile contributes an additional 4~\gcm\~.

\section{Comparison of Final Cut Data to Proton and Iron Simulations}\label{sec:TAPAO}
\begin{figure}[tbp!]
\centerline{
\includegraphics[width=0.9\linewidth]{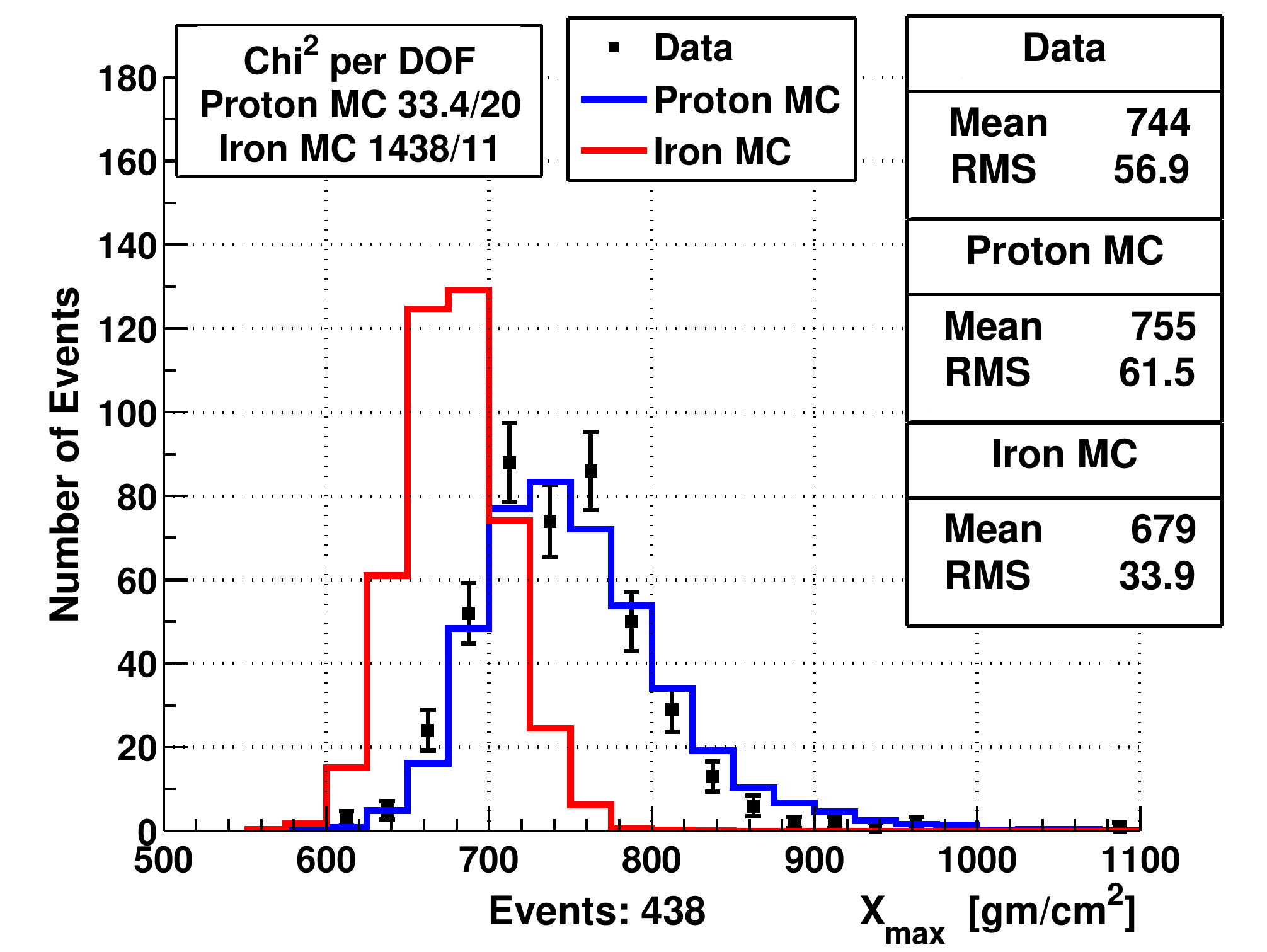}
}
\caption[Optimized pattern recognition data/MC comparisons of the shower maximum (\xmax): the distribution of measurements, for E~$>$10$^{18.2}$~eV, is shown for the data (black points with error bars) with the QGSJETII-03 proton MC (blue), and the iron MC (red) histograms.]{The data/MC comparisons of the shower maximum (\xmax), for E~$>$10$^{18.2}$~eV: the distribution of measurements is shown for the data (black points with error bars), the QGSJETII-03 proton MC (blue), and iron MC (red) histograms. The MC has been normalized to the area of the data. The binned maximum likelihood estimated chi-squared test values show much better agreement between data and proton. Data is not in agreement with iron.} 
\label{fig:AllDataMCComparison}
\end{figure}

\begin{figure}[tbp!]
  \centerline{
   \includegraphics[width=0.9\linewidth]{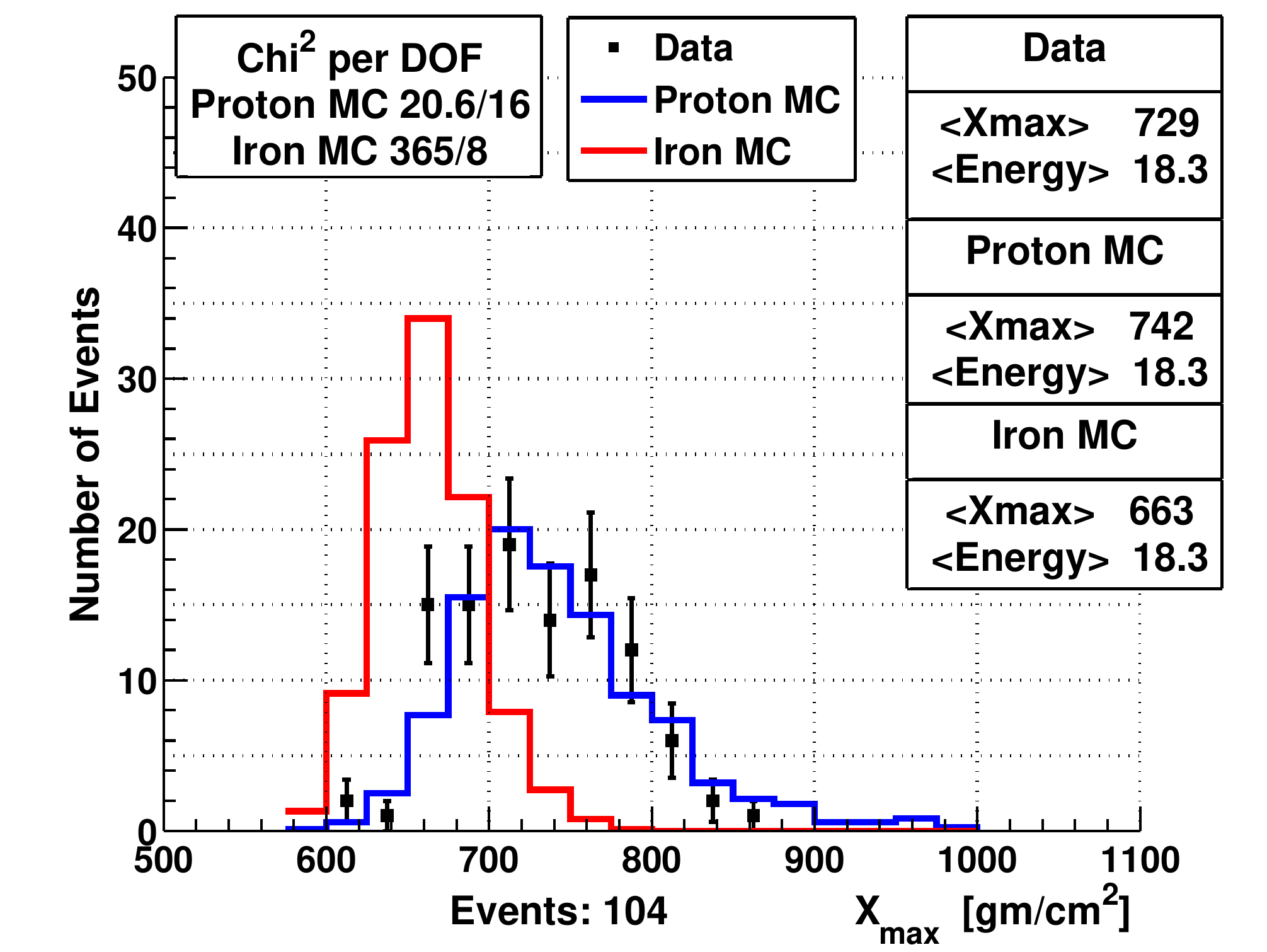}
   }
   \caption{The \xmax\ distributions from the data (black points), QGSJETII-03 proton MC (blue histogram), and iron MC (red histogram): energy range $=$ $18.20 < \log_{10} (E/eV) < 18.4$.} 
  \label{fig:184Xmax}
\end{figure}

\begin{figure}[tbp!]
  \centerline{
   \includegraphics[width=0.9\linewidth]{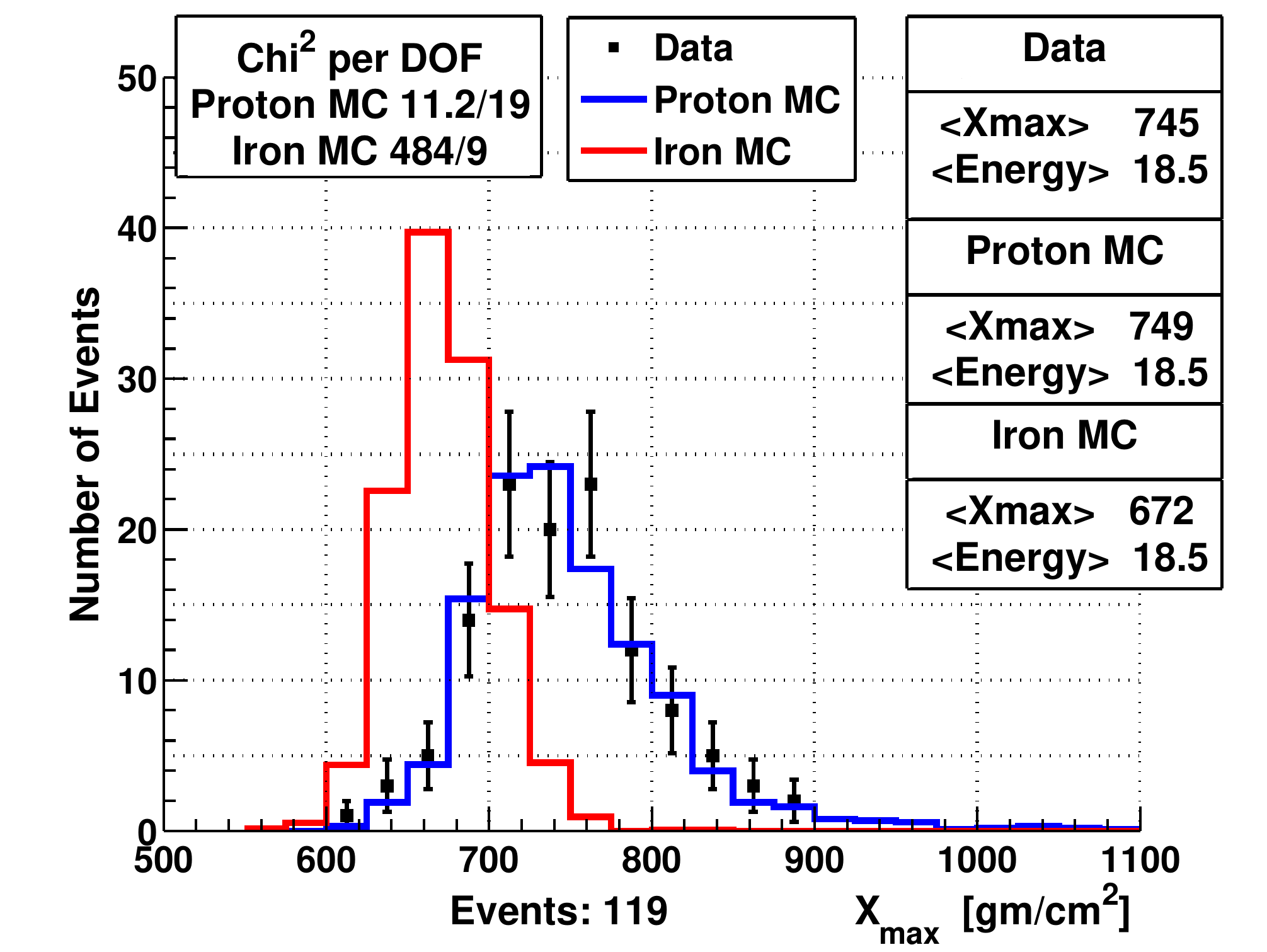}
   }
   \caption{The \xmax\ distributions from the data (black points), QGSJETII-03 proton MC (blue histogram), and iron MC (red histogram): energy range $=$ $18.4 < \log_{10} (E/eV) < 18.6$.} 
  \label{fig:186Xmax}
\end{figure}

\begin{figure}[tbp!]
  \centerline{
   \includegraphics[width=0.9\linewidth]{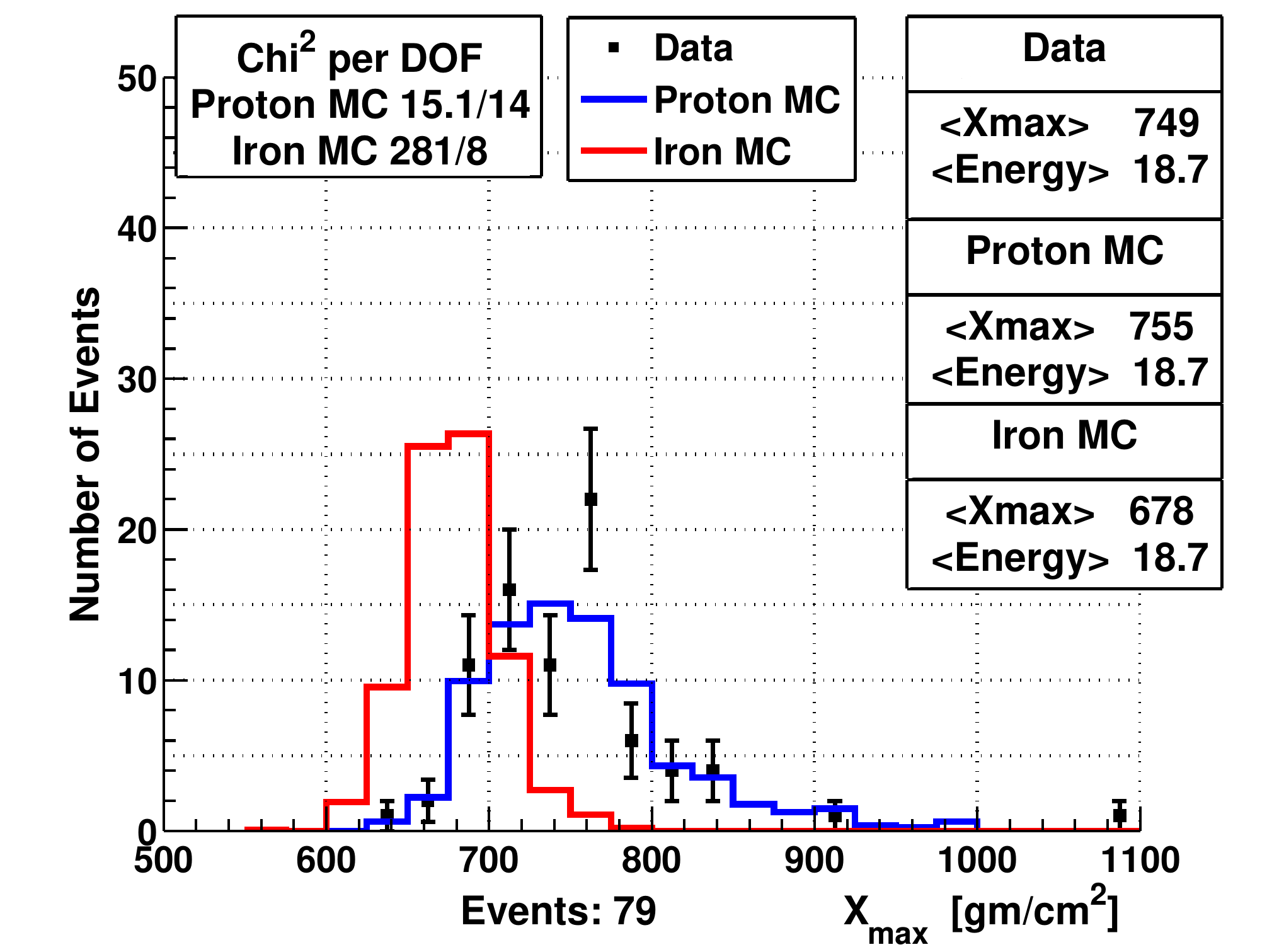}
   }
   \caption{The \xmax\ distributions from the data (black points), QGSJETII-03 proton MC (blue histogram), and iron MC (red histogram): energy range $=$ $18.6 < \log_{10} (E/eV) < 18.8$.} 
  \label{fig:188Xmax}
\end{figure}

\begin{figure}[tbp!]
  \centerline{
   \includegraphics[width=0.9\linewidth]{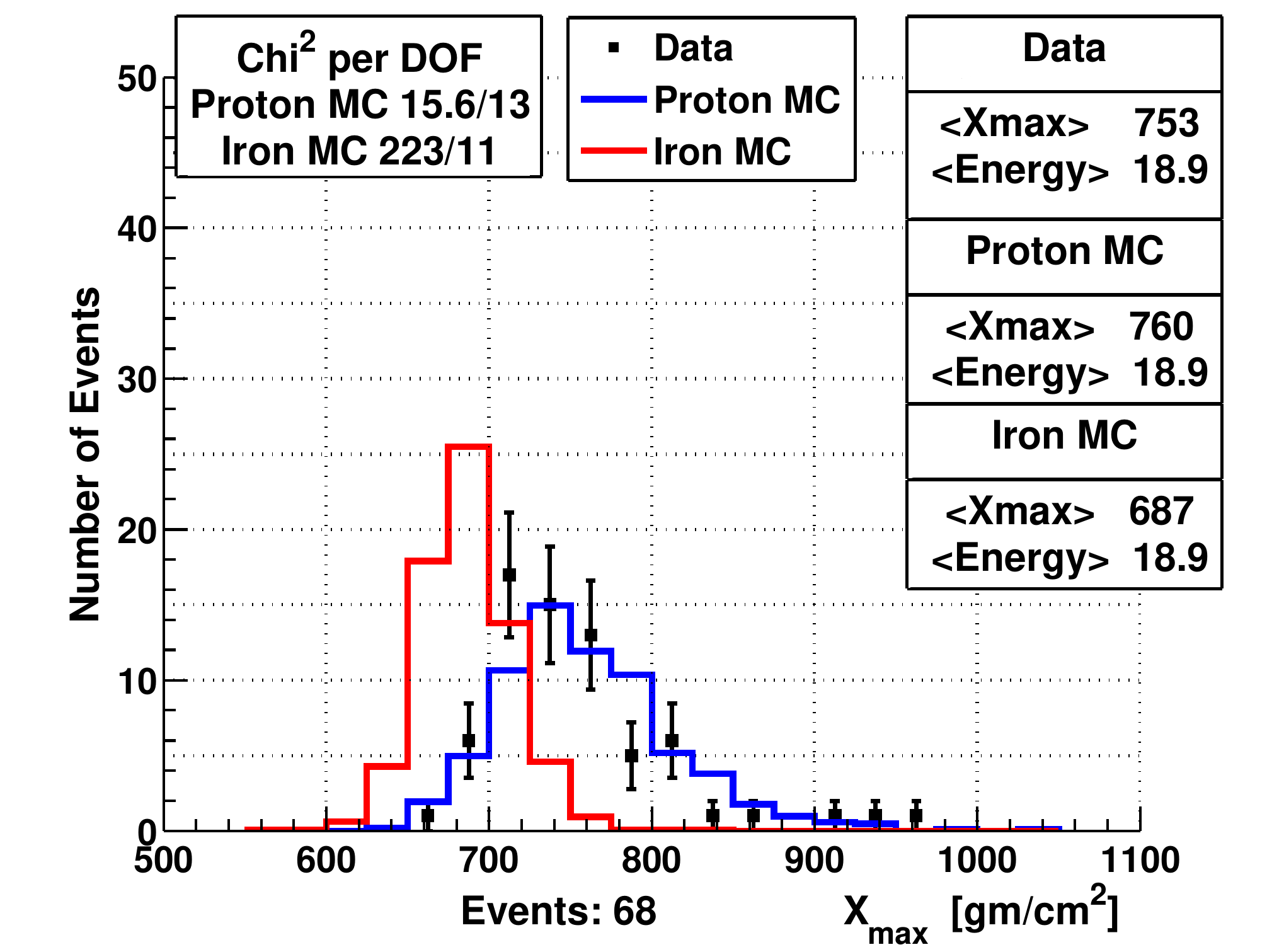}
   }
   \caption{The \xmax\ distributions from the data (black points), QGSJETII-03 proton MC (blue histogram), and iron MC (red histogram): energy range $=$ $18.8 < \log_{10} (E/eV) < 19.0$.} 
  \label{fig:19Xmax}
\end{figure}

\begin{figure}[tbp!]
  \centerline{
   \includegraphics[width=0.9\linewidth]{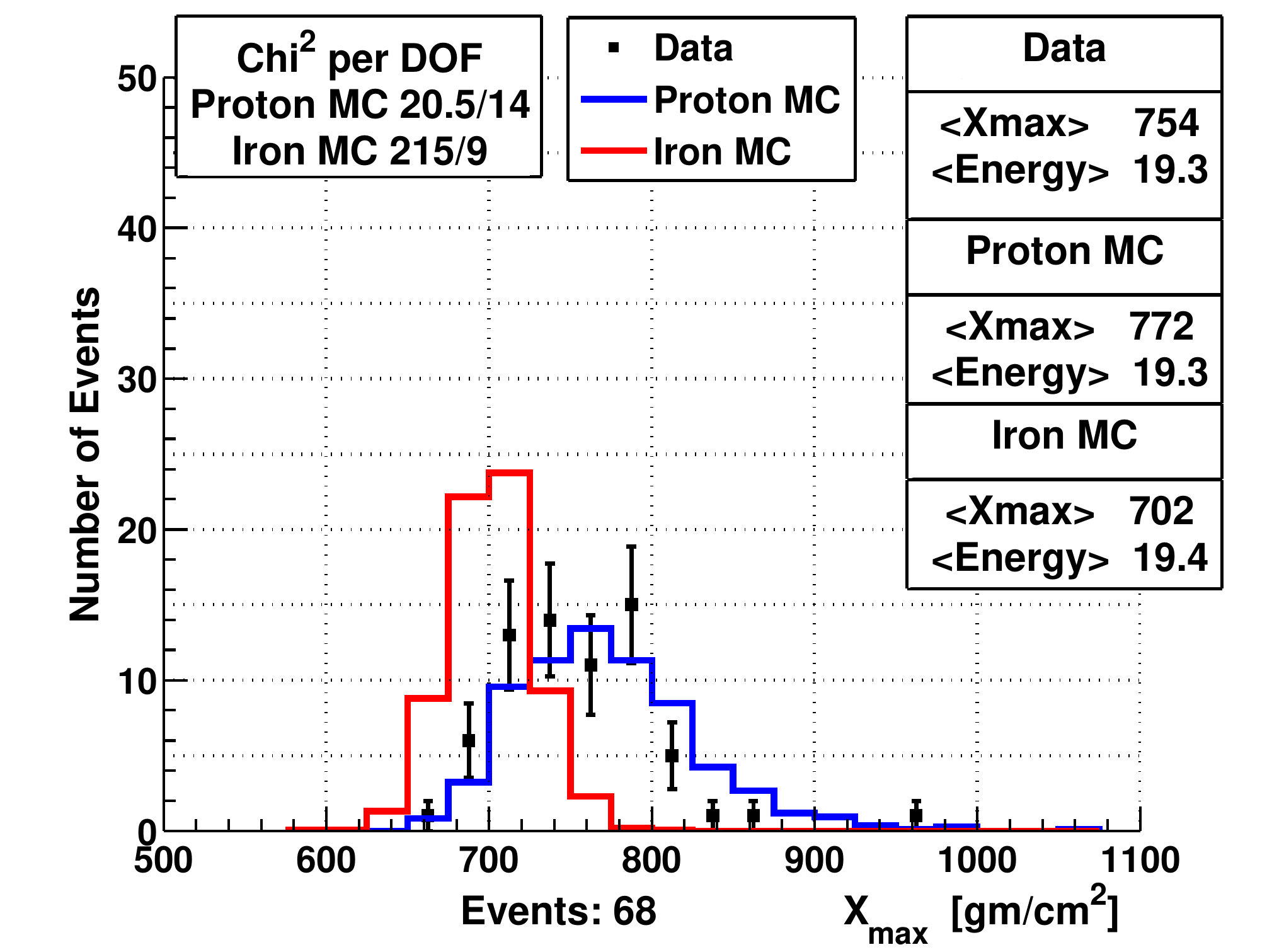}
   }
   \caption{The \xmax\ distributions from the data (black points), QGSJETII-03 proton MC (blue histogram), and iron MC (red histogram): energy range $=$ $ \log_{10} (E/eV) > 19$.} 
  \label{fig:Greater19Xmax}
\end{figure}

\begin{figure}[tbp!]
  \centerline{
   \includegraphics[width=1.0\linewidth]{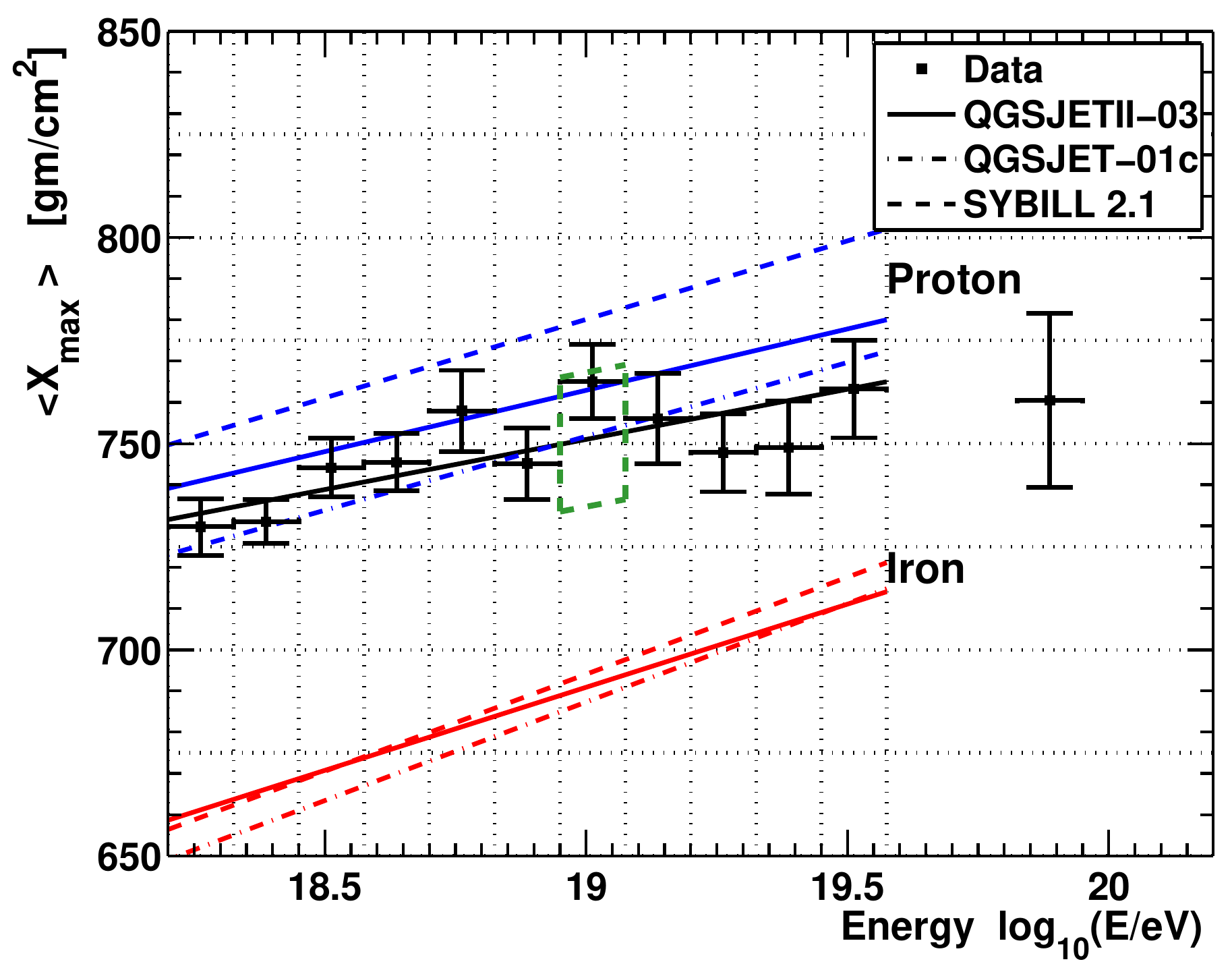}
   }
 \caption{The final Middle Drum hybrid composition result using geometry and pattern recognition cuts, for QGSJET-01c, QGSJETII-03, and SIBYLL 2.1 hadronic models. Data are the black points with error bars. The solid black line is a fit to the data. Colored lines are fits to MC. Blue is proton and red is iron. The green hashed box indicates the total systematic error on $<$\xmax$>$.}
  \label{fig:allrails}
\end{figure}

Figure \ref{fig:AllDataMCComparison} shows the \xmax\ distribution comparison, for optimized pattern recognition events, over the entire energy range (E~$>$10$^{18.2}$~eV).  Figures \ref{fig:184Xmax} - \ref{fig:Greater19Xmax} show the distributions in bins of width 0.2~in~$\log_{10}(E)$. There are at least 68 events in each bin. All bins with E~$>$10$^{19}$~eV are combined due to low statistics. For each energy bin the data is in good agreement with the proton MC. The binned maximum likelihood estimated chi-squared test values~\cite{Baker1984}, for each pair of distributions, are shown on each plot. The proton comparisons are in much better agreement, than iron, with the data over the entire energy range. This agreement extends over a variety of hadronic models, as far as the elongation rate is concerned (See Figure \ref{fig:allrails}). 

Note that, since the estimated systematic uncertainty (at Energy=10$^{19}$) of the mean \xmax\ is 16.3~\gcm\, and the statistical uncertainty resulting from the linear fit (as shown in Figure~\ref{fig:ERateTotalTri}) is 9.4~\gcm, both QGSJET-I-c and QGSJET-II-03 are in reasonable agreement with the data, for a light, largely protonic, composition. The SIBYLL~2.1 model~\cite{Sibyll2.1} for protons is 20-30~\gcm\ deeper than the data elongation rate. If the SIBYLL~2.1 model is correct, it would require an admixture of alpha particles, and CNO nuclei to the protons to describe the data precisely. More recent hadronic models are in progress. A recent monocular FD composition study shows that, when compared to SIBYLL~2.1, QGSJETII-04 is only $\sim$2~\gcm\ shallower, and EPOS-LHC is expected to give a 20~\gcm\ deeper \xmax\ result~\cite{Fujii2014}.

The PAO results indicate an RMS narrowing of the \xmax\ distribution relative to expectations for protons, at energies greater than 10$^{18.5}$~eV. At the current level of statistics this paper cannot support, or rule out, such an effect because of statistical sampling bias, particularly at the highest energies. Definitive statements about this claim await the completed analysis of additional hybrid data from the Black~Rock and Long~Ridge fluorescence detector sites, as well as purely stereo data from all three sites. 

\section{Conclusion}\label{sec:conclusion}
The importance of this paper is in its use of fluorescence detectors, identical to HiRes, with a hybrid reconstruction technique. The HiRes composition result used a stereo reconstruction method, while this paper uses a hybrid technique, similar but not identical, to one used by the PAO group. It is therefore important that the current hybrid TA data is in good agreement with the HiRes results, as this indicates that differences in aperture, reconstruction, and modeling by Monte Carlo simulations do not lead to any significant systematic differences in the final physics result in the case of identical fluorescence detectors.

The measured average \xmax\ at 10$^{19}$~eV is \\751$\pm$16.3~sys.$\pm$9.4~stat.~\gcm\ and the elongation rate is 24.3$\pm$3.8~sys.$\pm$6.5~stat.~\gcm.  Assuming a purely protonic composition, taking into account all reconstruction and acceptance biases (using the QGSJETII-03 model), we would expect the average \xmax\ at~10$^{19}$ eV to be 763~\gcm\, and the elongation rate to be~29.7~\gcm\ per energy decade.

Considering the fact that TA hybrid, and PAO hybrid data, have different acceptances, and analysis techniques, a direct comparison of the results can be misleading. Detailed comparisons, using a set of simulated events from a mix of elements that are in good agreement with the PAO data, are in progress~\cite{Array:2013dra}. Such a mix can be input into the TA hybrid simulation, and reconstruction programs, and the result will be a prediction of what TA should observe given a composition inferred from PAO data. A direct comparison with the TA data can then be made. Since this work is in progress, we simply remark that a light, nearly protonic, composition is in good agreement with the data, for both simple geometric cuts and pattern recognition cuts that result in improved \xmax\ resolution.

\section{Acknowledgements}
The Telescope Array experiment is supported by the Japan Society for
the Promotion of Science through Grants-in-Aids for Scientific
Research on Specially Promoted Research (21000002) ``Extreme Phenomena
in the Universe Explored by Highest Energy Cosmic Rays'' and for
Scientific Research (19104006), and the Inter-University Research
Program of the Institute for Cosmic Ray Research; by the U.S. National
Science Foundation awards PHY-0307098, PHY-0601915, PHY-0649681,
PHY-0703893, PHY-0758342, PHY-0848320, PHY-1069280, and PHY-1069286;
by the National Research Foundation of Korea (2007-0093860, R32-10130,
2012R1A1A2008381, 2013004883); by the Russian Academy of Sciences,
RFBR grants 11-02-01528a and 13-02-01311a (INR), IISN project
No. 4.4509.10 and Belgian Science Policy under IUAP VII/37 (ULB). 
This work was supported by the World Premier International Research 
Center Initiative (WPI Initiative), MEXT, Japan. 
The foundations of Dr. Ezekiel R. and Edna Wattis Dumke, Willard L. Eccles
and the George S. and Dolores Dore Eccles all helped with generous
donations. The State of Utah supported the project through its
Economic Development Board, and the University of Utah through the
Office of the Vice President for Research. The experimental site
became available through the cooperation of the Utah School and
Institutional Trust Lands Administration (SITLA), U.S. Bureau of Land
Management, and the U.S. Air Force. We also wish to thank the people
and the officials of Millard County, Utah for their steadfast and warm
support. We gratefully acknowledge the contributions from the
technical staffs of our home institutions. An allocation of computer
time from the Center for High Performance Computing at the University
of Utah is gratefully acknowledged.


\newpage

\bibliographystyle{elsarticle-num}
\bibliography{library}

\end{document}